\documentclass[useAMS,usenatbib]{mn2e}
\title[Magnetic fields and the dynamics of spiral galaxies]
{Magnetic fields and the dynamics of spiral galaxies}
\author[C. L. Dobbs \& D.  J. Price]
{C. L. Dobbs\thanks{E-mail:
dobbs@astro.ex.ac.uk} \& D. J. Price \\
School of Physics, University of Exeter, 
Stocker Road, Exeter, EX4 4QL \\}
\usepackage{amssymb}
\usepackage{amsmath}
\usepackage{graphicx}

\begin{document}
\date{\today}

\pagerange{\pageref{firstpage}--\pageref{lastpage}} \pubyear{0000}

\maketitle

\label{firstpage}

\begin{abstract}
We investigate the dynamics of magnetic fields in spiral galaxies by performing 3D Magnetohydrodynamics (MHD) simulations of galactic discs subject to a spiral potential using cold gas, warm gas and a two phase mixture of both. Recent hydrodynamic simulations have demonstrated the formation of inter-arm spurs as well as spiral arm molecular clouds provided the ISM model includes a cold HI phase. We find that the main effect of adding a magnetic field to these calculations is to inhibit the formation of structure in the disc. However, provided a cold phase is included, spurs and spiral arm clumps are still present if $\beta \gtrsim 0.1$ in the cold gas. A caveat to the two phase calculations though is that by assuming a uniform initial distribution,  $\beta \gtrsim 10$ in the warm gas, emphasizing that models with more consistent initial conditions and thermodynamics are required. Our simulations with only warm gas do not show such structure, irrespective of the magnetic field strength.

Furthermore, we find that the introduction of a cold HI phase naturally produces the observed degree of disorder in the magnetic field, which is again absent from simulations using only warm gas. Whilst the global magnetic field follows the large scale gas flow, the magnetic field also contains a substantial random component that is produced by the velocity dispersion induced in the cold gas during the passage through a spiral shock.
Without any cold gas, the magnetic field in the warm phase remains relatively well ordered apart from becoming compressed in the spiral shocks.
Our results provide a natural explanation for the observed high proportions of disordered magnetic field in spiral galaxies and we thus predict that the relative strengths of the random and ordered components of the magnetic field observed in spiral galaxies will depend on the dynamics of spiral shocks.
\end{abstract}


\section{Introduction}
Observations of the magnetic field in galaxies indicate that the magnetic pressure is comparable in magnitude to the thermal pressure and interstellar turbulence \citep{Heiles2005}. Consequently magnetic fields are expected to play an important part in the dynamics of the ISM, including spiral shocks and the formation of molecular clouds.
  
The response of a gaseous disc to a spiral potential has been well examined both theoretically and numerically \citep*{Roberts1969,Shu1972,Woodward1976,Gittins2004,Wada2004,DBP2006}. 
The gas flow evolves into a quasi-stationary solution which contains spiral shocks, providing the potential is of sufficient strength. \citet{Roberts1970} found that the strength of the shock decreases with increasing magnetic field strength, whilst the magnetic field strength is amplified in the spiral shock, the latter agreeing with observations at the time for galactic magnetic fields.

There are numerous hydrodynamic simulations of gas discs subject to a spiral potential (e.g. \citealt{Kim2002,Chak2003,Slyz2003,Wada2004,Gittins2004,Dobbs2006}) which discuss the formation of substructure and location of the shock. Whilst 2D simulations of warm gas show the formation of small-scale spurs perpendicular to the spiral arms \citep{Gittins2004,Wada2004}, these features appear to be suppressed in 3D simulations \citep{Kim2006}. On the other hand, \citet{Dobbs2006} do find more extensive substructure and spiral arm clouds \citep*{DBP2006,Dobbs2007} in 3D simulations, but only in models where the gas temperature is $\leq 1000$ K. There are comparatively fewer MHD calculations, largely due to the relatively recent addition of magnetic fields into numerical codes, and the limitations of numerical resolution. 
\citet{Shetty2006} investigate the formation of spurs using 2D MHD grid-based calculations of the warm ISM with self gravity, concluding that self gravity is necessary for spurs to evolve.
\citet{Gomez2002,Gomez2004} also perform 3D calculations with warm gas, but whilst both groups compare MHD with non-MHD results, neither investigate the effects of varying the field strength or ISM temperature. 

Recent observations now provide much more detailed measurements of the magnetic field strength and direction in the diffuse (warm) ISM (see reviews by \citealt{Beck1996} and \citealt{Beck2007}). 
Spiral magnetic arms appear to occur in all disc galaxies, regardless of the presence of optical spiral arms \citep{Beck2005}. 
Typically the field in the inter-arm regions is several $\mu$G, but may be 10 or more $\mu$G in the spiral arms, and the total field contains random and ordered (regular) components of comparable strength.
The general consensus currently favours dynamo theory to explain the origin of these relatively strong magnetic fields \citep*{Parker1971,P1971,Balsara2004,Beck2005}, whereby the magnetic field is generated by turbulence in the ISM, and can produce or enhance magnetic spiral arm arms \citep*{Panesar1992,Rohde1998}. However in grand design galaxies where the gas experiences strong spiral shocks, such as M51, the magnetic spiral arms tend to be aligned with the dust lanes \citep{Nein1992} suggesting that the magnetic field is strongly related to the dynamics of the spiral shock. 

Magnetic fields have been long associated with molecular cloud formation, originally through the Parker instability \citep{Parker1966}, and more recently via the magneto-rotational 
\citep*{Kim2003,Piontek2005} instability. The Parker instability is expected to produce sinusoidal motions in the $z$ direction, with the magnetic field channeling gas into dense concentrations in the plane of the disc. Molecular clouds formed in this way were originally believed to be magnetically supported, with lifetimes of order $10^8$ years \citep*{Zweibel1975,Shu1987}.
However the Parker instability is found to be relatively weak, and singularly insufficient to induce molecular cloud formation \citep*{Elmegreen1982,OtherKim2001,KOS2002}. The magneto-rotational instability (MRI) has been proposed as a formation mechanism away from spiral arms \citep{Kim2003}. However the MRI generally takes a few orbits to initialise, particularly if the field is toroidal (e.g. 10's of orbits in simulations by \citealt{Nish2006}), and is therefore of less importance where spiral shocks occur.
Magnetic fields may be more relevant to GMC (giant molecular cloud) formation in conjunction with gravitational instabilities \citep{ElmegreenMJI1987,Kim2001,KOS2002} [also known as the magneto-Jeans-instability, MJI] or cooling \citep{Kosinski2007}.
 
In this paper we describe 3D MHD calculations of a galactic disc subject to a spiral perturbation. These are the first fully Lagrangian MHD calculations to model this problem, using the Smoothed Particle Magneto-hydrodynamics (SPMHD) code \citep{Price2004,Price2005,PB2007}.
We compare the structure of the disc for a range of magnetic field strengths, assuming a single phase medium of cold or warm gas, or a two-phase medium. 
We also describe the strength and morphology of the magnetic field, and relate these to observations. 
In this paper we focus on the effect of the spiral shock inducing structure in the gas rather than Parker or MRI instabilities, and we leave a discussion of results including self-gravity to a future paper. 
 
\section{Calculations}
We use a 3D Smoothed Particle Magnetohydrodynamics code (SPMHD) for these calculations, based on a binary tree code developed by Benz \citep{Benz1990}. The code has been extensively modified, including the use of sink particles \citep{Bate1995}, and in particular, the addition of magnetic fields \citep{Price2004,Price2004b,PB2007}.

The SPMHD code solves the equations of magnetohydrodynamics in the form
\begin{eqnarray}
\frac{d\rho}{dt} & = & -\rho \frac{\partial v^i}{\partial x^i}, \label
{eq:cty} \\
\frac{dv^i}{dt} & = & \frac{1}{\rho}\frac{\partial S^{ij}}{\partial
x^j} + f^{i}_{ext}, \label{eq:mom} \\
\frac{du^i}{dt} & = & -\frac{P}{\rho}\frac{\partial v^{i}}{\partial
x^i}, \label{eq:ener} \\
{\bf B} & = & \nabla\alpha_{E} \times \nabla \beta_{E}, \label{eq:ind1} \\
\frac{d\alpha_{E}}{dt} & = & 0, \hspace{1cm} \frac{d\beta_{E}}{dt} = 0, \label{eq:ind2}
\end{eqnarray}
where $\rho$, $v^{i}$, $P$, $u$, $P$ and ${\bf B}$ are the fluid
density, velocity, pressure, thermal energy per unit mass and
magnetic flux respectively, $\frac{d}{dt} \equiv \frac{\partial}
{\partial t} + v^i \frac{\partial}{\partial x^i}$ is the time
derivative following the motion of a fluid particle and $S^{ij}$ is
the stress tensor given by
\begin{equation}
S^{ij} = -P \delta^{ij}  + \frac{1}{\mu_0} \left( B^i B^j- \frac12
\delta^{ij}
B^2 \right).
\end{equation}
The numerical representation of these equations as summations over
SPH particles is discussed in detail in \citet{Price2004,Price2004b,Price2005} and \citet{PB2007}. In particular equations (\ref{eq:ind1})-(\ref{eq:ind2}) use the `Euler potentials' representation for the magnetic field \citep{PB2007,Stern1970}, for which the Lagrangian evolution is zero (Equation \ref{eq:ind2}), corresponding to the advection of magnetic field lines in ideal MHD. The advantage of using this formulation is that the divergence constraint on the magnetic field is satisfied by construction.

Dissipative terms in the form of an artificial viscosity and
resistivity are added to the momentum and induction equations on
order to correctly capture discontinuities in the flow (i.e. shocks
and magnetic current sheets).
Artificial viscosity is used with the standard parameters $\alpha_V=1$
and $\beta_V=2$ \citep{Monaghan1997,Price2005}, as developed for
the SPMHD code and artificial resistivity is added to the Euler potentials' evolution with $\alpha_B=1$ as described in \citet{PB2007}.

A galactic potential is added via an external driving force $f_{ext}
$, the exact form of which is described in \citet{Dobbs2006}. The
main features of the spiral component (from \citealt{Cox2002}) are a
4 armed spiral pattern, of pattern speed
$2 \times 10^8$ rad yr$^{-1}$ and pitch angle $15^o$. The amplitude of the spiral potential is $1.1 \times 10^{12}$ cm$^2$ s$^{-2}$. The total potential also includes components for the disc and halo.

The code also includes individual time-steps, and variable smoothing
lengths. The calculation of the smoothing lengths uses a recent
improvement by which the density and smoothing are solved iteratively
for each particle \citep{Price2004,PM2007}. The density and smoothing length
are related according to
\begin{equation}
h=\eta \bigg(\frac{m}{\rho}\bigg)^{1/3}
\end{equation}
where $h$ is the smoothing length, $\rho$ the density, $m$ the mass
of the particle and $\eta$ is a dimensionless parameter set to 1.2 in
order that each particle has $\sim$ 60 neighbours. This gives a mean value of $h\sim 30$ pc initially (40 pc in the multi-phase simulations) using 4 million particles, such that $h/H\sim0.2$. Although the scale height ($H$) of the cold gas decreases (see next section) with time, the density of the cold component increases considerably in the spiral arms, so $h$ also decreases in the arms. Overall the $z$ dimension tends to be marginally resolved in the spiral arms but not well resolved in the inter-arm regions.   

 \subsection{Initial distribution of gas}   
We place particles in a torus with radii 5 kpc $<r<$ 10 kpc. The gas is initially distributed randomly  and the velocities are assigned with circular orbits. Unlike previous results \citep{Dobbs2006}, where the initial positions and velocities were extracted from a test calculation, the spiral perturbation emerges with time in these simulations.
The velocities in the plane of the disc follow a rotation curve corresponding to the disc component of the potential
\begin{equation}
\psi_{disc}=\frac{1}{2} v_0^2 \: log \bigg(r^2+R_c^2+(z/z_q)^2\bigg)
\end{equation} 
where $R_c$=1 kpc, $v_0=220$ km s$^{-1}$, and $z_q$=0.7 is a measure of the disc scale height.
This produces an essentially flat rotation curve for the radii over which the particles are distributed.
The velocities in the vertical direction are chosen from a Gaussian of mean 6 km s$^{-1}$.
The total mass in the torus is $10^9$ M$_{\odot}$, giving an initial surface density of 4 M$_{\odot}$pc$^{-2}$, similar to that of HI near the solar radius \citep*{Wolfire2003}. In all these calculations we use 4 million particles.

We perform single phase and two phase calculations. The gas is either cold (100 K) or warm ($10^4$~K) in the single phase models. For the two phase calculations, we use equal proportions of warm and cold gas. Each phase contains 2 million particles placed in a uniform random distribution over the disc. As the simulation progresses 
the distribution evolves into cold clumps surrounded by a warm tenuous medium \citep{Dobbs2007}.
All calculations are isothermal, and we do not include self gravity or feedback from star formation. The assumption of an isothermal gas is not ideal, and for the two-phase simulations, we would expect heating and cooling between the two components. Our analysis instead assumes that there is a reservoir of cold HI in the ISM as well as warm gas \citep{Heiles2003,Gibson2006}, and we consider the response of such a distribution to a spiral density wave. 

In the single phase simulations, we take an initial maximum height of 150 kpc, whilst for the two phase simulations the initial maximum height is 400~pc. These values were chosen based on the observed scale heights for the Galaxy \citep{Cox2005}, for cold and warm gas. However assuming pressure equilibrium, and that the density distribution corresponding to Eqn.~8 falls off as $z^{-2}$ \citep{Binney}, the scale height expected from the logarithmic potential is:
\begin{equation}
\bigg(\frac{H}{z_q}\bigg)^2=2 \bigg(\frac{c_s}{v_0} \bigg)^2 (r^2+R_c^2+(z/z_q)^2) \approx 110  \bigg(\frac{c_s}{v_0} \bigg)^2
\end{equation} 
at a radius of 7.5 kpc, where $c_s$ is the sound speed of the gas.
This leads to a scale heights of approximately 33 and 330 pc in the cold and warm phases respectively at a radius of 7.5 kpc.  During the simulation, the scale height of the warm gas settles into an exponential distribution of scale height 300-340 pc. 
We note that without supernovae, a lower scale height for the warm gas is expected than implied by observations ($\sim130$ pc \citep{McKee1990}), in which case a lower value of $z_q$ may be appropriate (0.25-0.3).

The scale height of the cold gas is initially maintained by the velocity dispersion applied in the $z$ direction. However the increase in density in the spiral arms and decay of the velocity dispersion means that the scale height of the cold gas is typically 20-100 pc at later times in both the single and two phase calculations. 
In the two phase simulations we did not assign a different initial height for the cold and warm components. However the results are not particularly sensitive to the initial scale height and do not change significantly if we set the initial scale height of the cold gas to a 
lower value.

The initial configuration of the magnetic field is toroidal, similar to previous calculations \citep{Shetty2006,Gomez2004}. A toroidal field can be described in terms of Euler potentials by
\begin{eqnarray}
\alpha_E & = & -B_0 \theta \\
\beta_E & = & \frac12 r^2
\end{eqnarray}
where $r^2 = x^2 + y^2 + z^2$ and $\theta = \cos^{-1}(z/r)$. The relative strength of the magnetic field is given by the mean plasma beta, defined as the ratio of gas to magnetic pressure\footnote[1]{The reader should note that in some papers (e.g. \citealt{Shetty2006}), $\beta$ is defined as half this ratio in which case $v_A=c_s/\sqrt{\beta}$.}
\begin{equation}
\beta= \frac{P}{B^2 /2 \mu_0}=\frac{2 \mu_0 \rho_0 c^2}{|B|^2},
\end{equation}
where $\rho_0$ is the average density of the disc. The Alfven speed is then $v_A=
B/\sqrt{\mu_0 \rho_0} = \sqrt{2} c_s/\sqrt{\beta}$.
With present observations of the diffuse (non-self-gravitating) ISM \citep{Heiles2005}, the magnetic pressure is estimated to be 3 times larger than the thermal component. The CNM and WNM are both found to exhibit both magnetic field strengths of $\sim 6 \mu$ G and a thermal pressure of $\sim 10400$ cm$^{-3}$ K \citep{HT2005} indicating that $\beta$ will be similar for each component. 
However in the two-phase simulations described in this paper (Section~3.4), we assume initial conditions of a uniform density disc of uniform field strength, so $\beta$ for the warm gas is 100 times that of the cold (that is, the field is relatively weaker for the warm gas). An alternative approach to be explored in future calculations is to set up the cold gas in dense clumps such that $\beta$ is uniform across the disc. 

\section{Results}
In this section we describe the structure of the disc for the single and multi-phase simulations. We ran these simulations for at least 300 Myr, by which time the structure of the spiral arms and morphology of the field were largely consistent with time. 
We discuss the morphology of the magnetic field, and the magnetic field strengths of the spiral arm and inter-arm regions in Sections~3.2-3.4. In Section~3.5 we compare the magnetic fields in our simulations with observations. 
\begin{figure*}
\centerline{
\includegraphics[scale=0.38,angle=270]{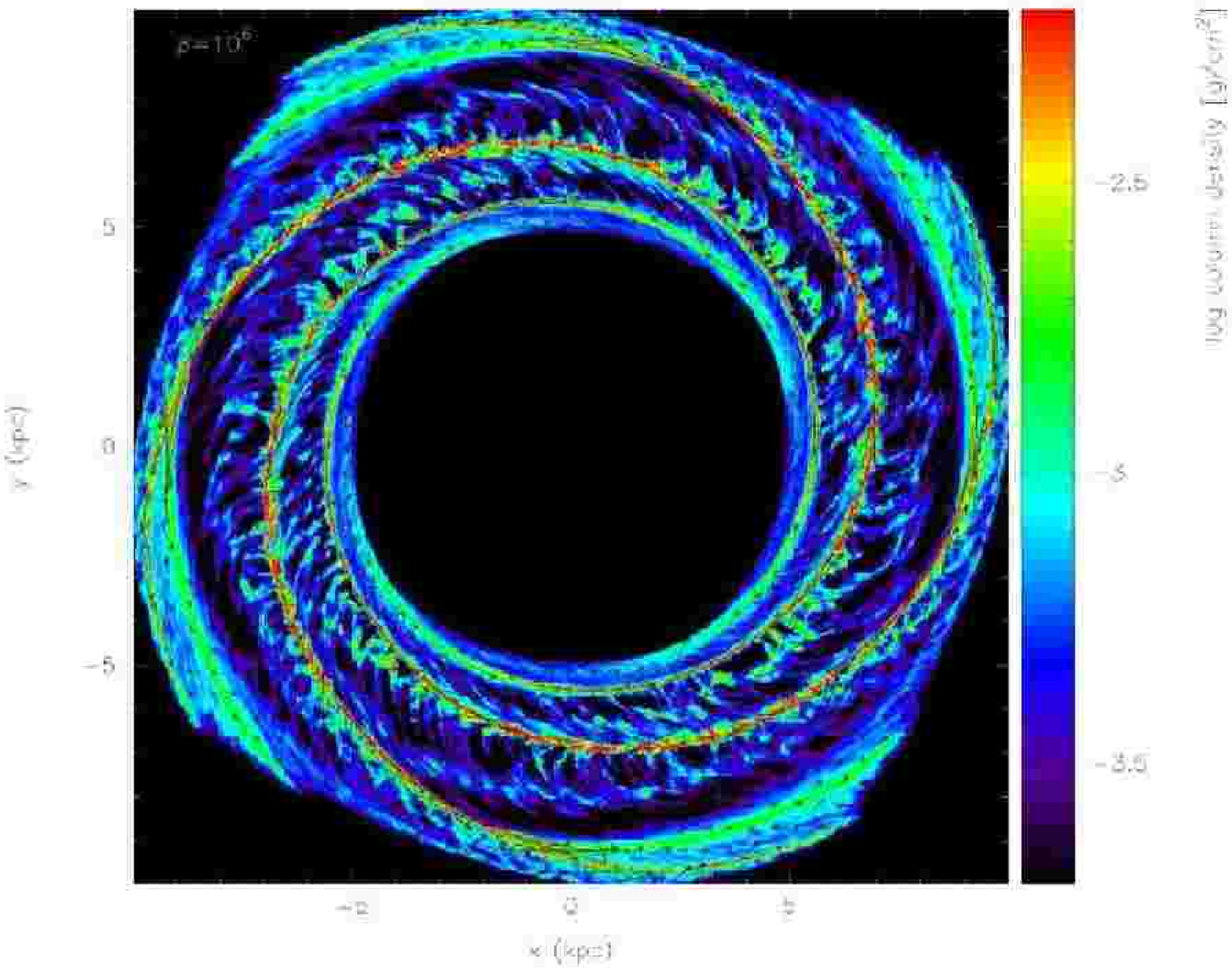}
\includegraphics[scale=0.38,angle=270]{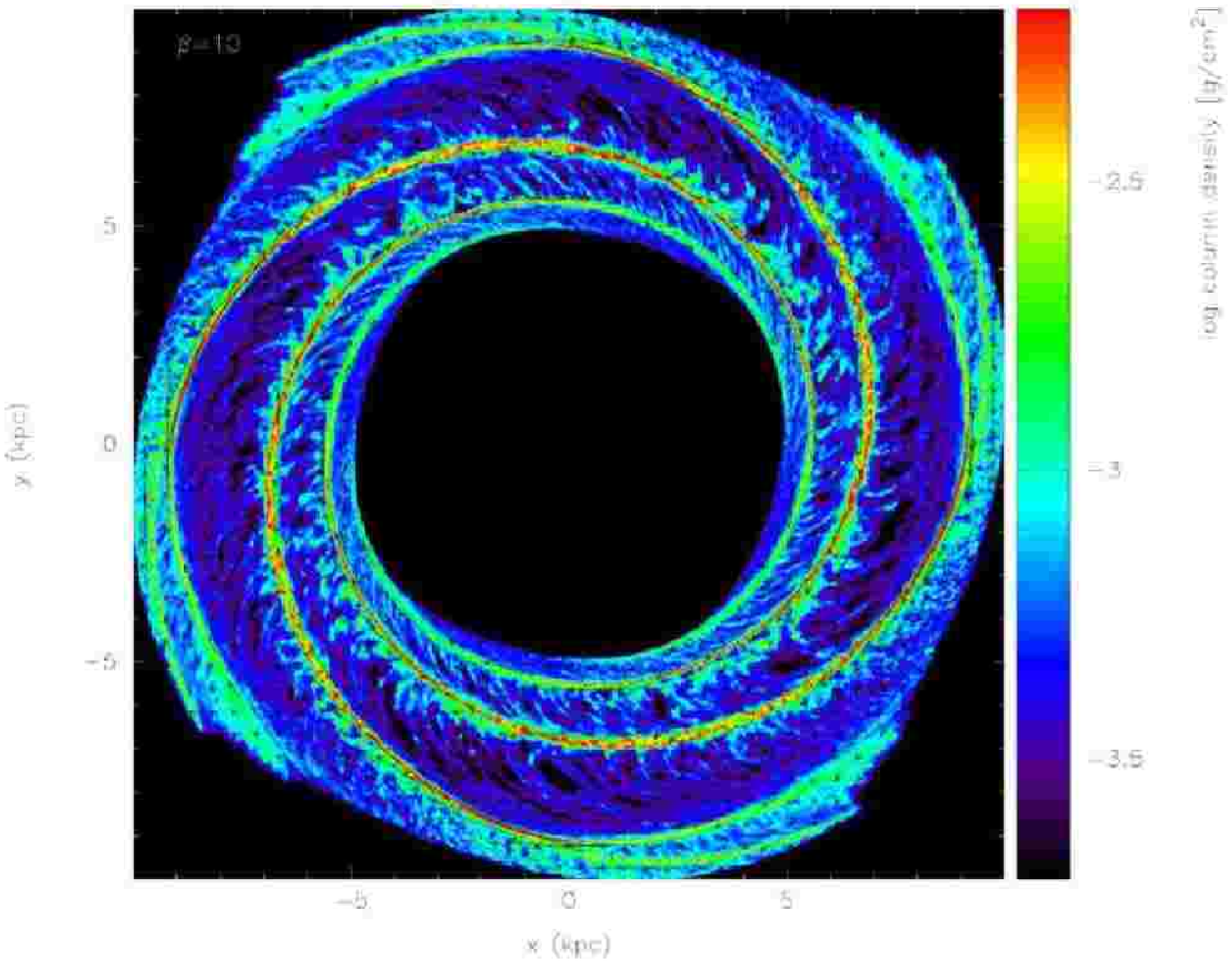}}   
\centerline{
\includegraphics[scale=0.38,angle=270]{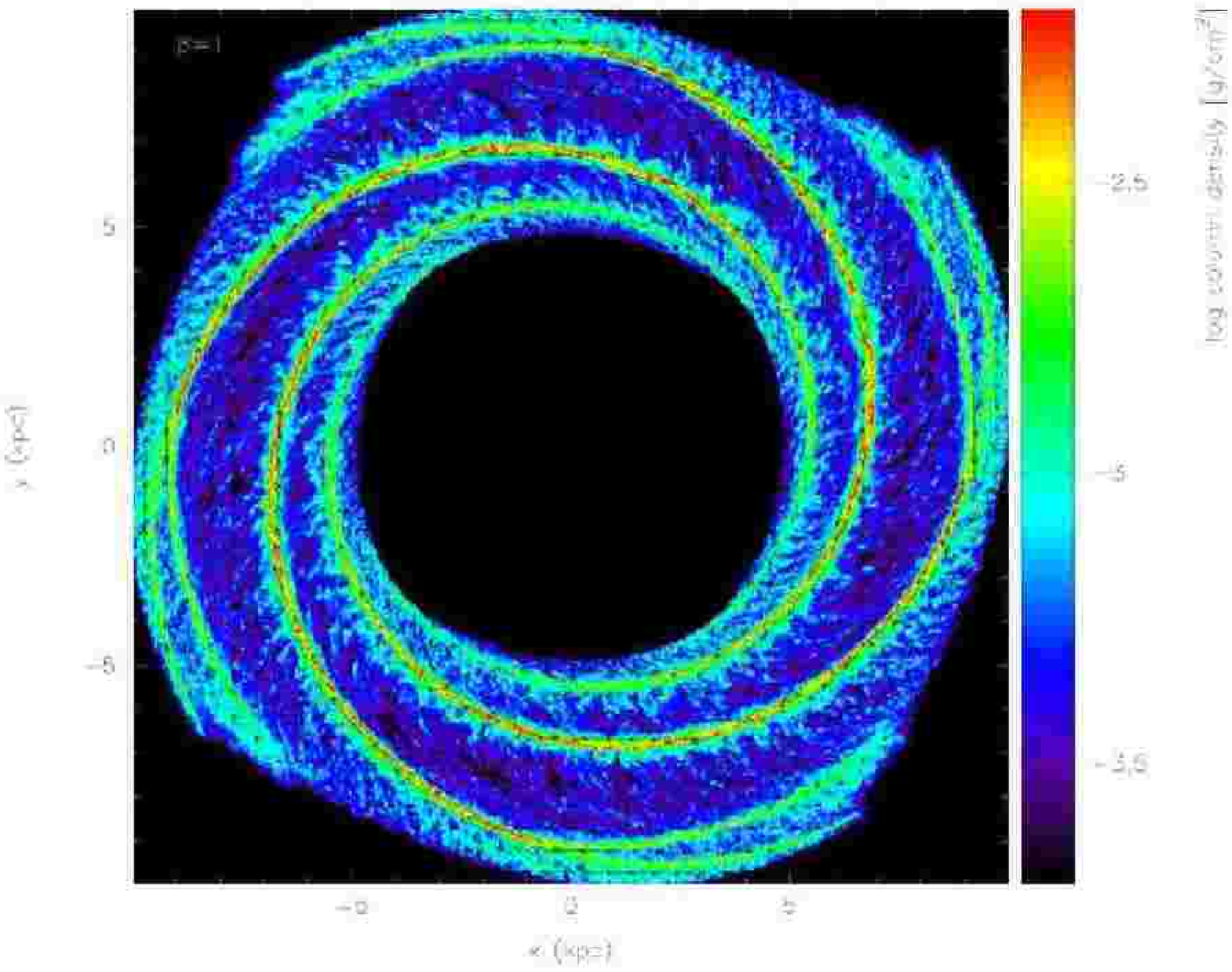}
\includegraphics[scale=0.38,angle=270]{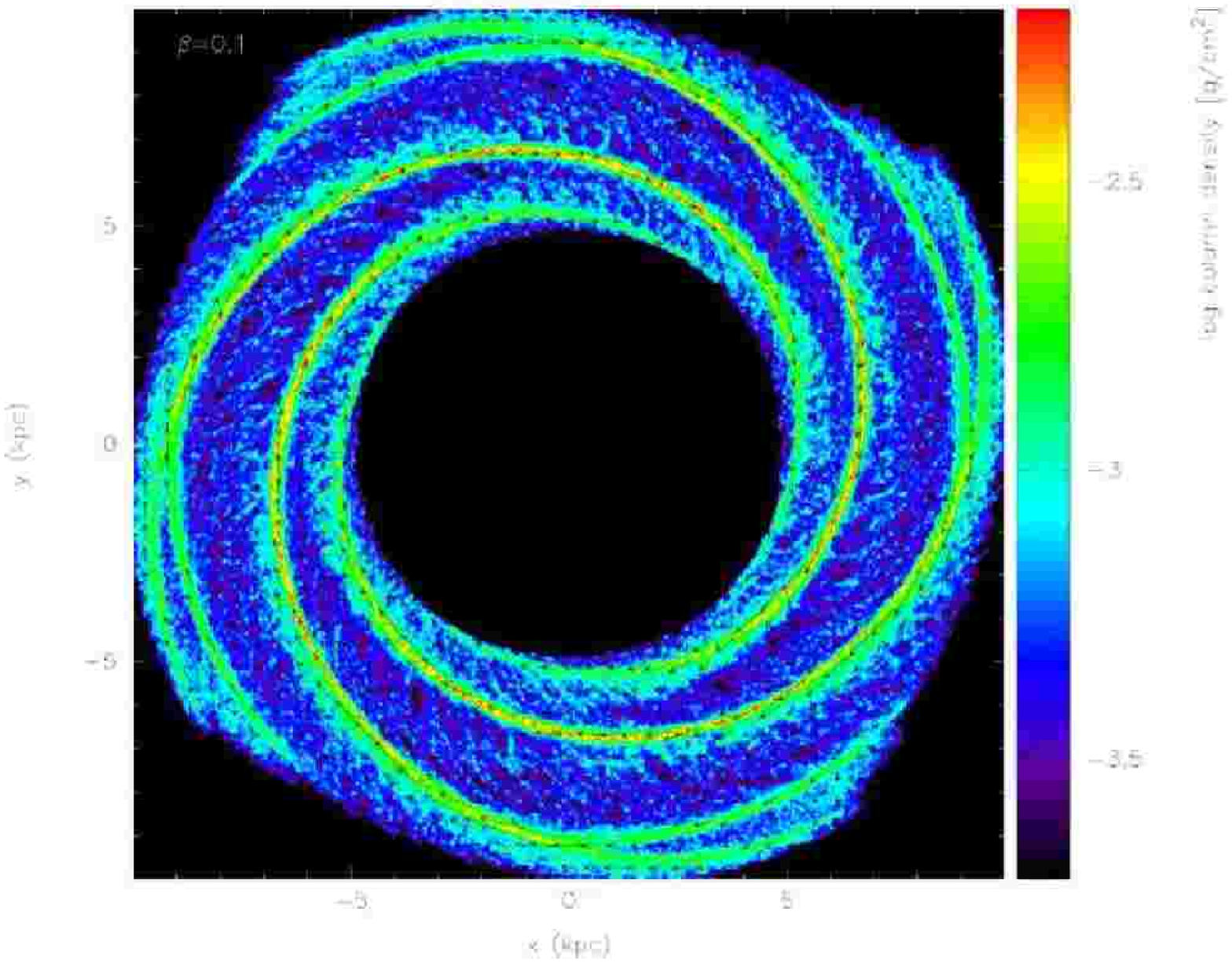}}
\caption{The spiral arm structure of the disc is shown for the single phase simulations with cold (100 K) gas after 250 Myr. The vectors indicate the magnetic field integrated through $z$, i.e. $\int B_x dz$ and $\int B_y dz$. The ratio of the thermal to magnetic pressure, $\beta$, for each plot (also marked) is $10^6$ (top left), 10 (top right), 1 (bottom left) and 0.1 (bottom right). The degree of structure in the disc is reduced as the magnetic field strength increases.}
\end{figure*}

\subsection{Single phase simulations}
We performed a series of simulations of increasing field strength, taking a gas temperature of 100 K. In addition, we ran two calculations with warm ($10^4$~K) gas. The parameters for these models are listed in Table~1.
Both the density and magnetic field strength increase during the simulations due to compression by shocks. The parameter $\beta$ also changes and as described in Section~3.2.4, much of the gas in the disc exhibits lower values of $\beta$ than indicated in Table~1. Furthermore the calculations are scale free, in that the density can be increased, corresponding to a square root increase in the magnetic field strength, and the structure of the disc is the same. 

\begin{table}
\centering
\begin{tabular}{c|c|c|c|c|c|c|c|c|c}
 \hline
 Model & T (K) & $\beta$ & B ($\mu$G) & $v_A$  (km s$^{-1}$)\\
 \hline
A & 100 & $10^6$ &  $3 \times 10^{-4}$ & $1.2 \times 10^{-3}$  \\
B & 100 & 10 & 0.09 & 0.35  \\
C & 100 & 1 & 0.3 & 1.2  \\
D & 100 & 0.1 & 0.9 & 3.5  \\ \hline
E & $10^4$ & 100 & 0.3 &  1.2 \\
F & $10^4$ & 1 & 3 & 12  \\
\hline
\end{tabular}
\caption{Table showing the parameters used in the single phase calculations. The magnetic field strengths correspond to a uniform initial ISM density of $\sim 0.7$ cm$^{-3}$.}
\end{table}
\begin{figure*}
\centerline{
\includegraphics[scale=0.38,angle=270]{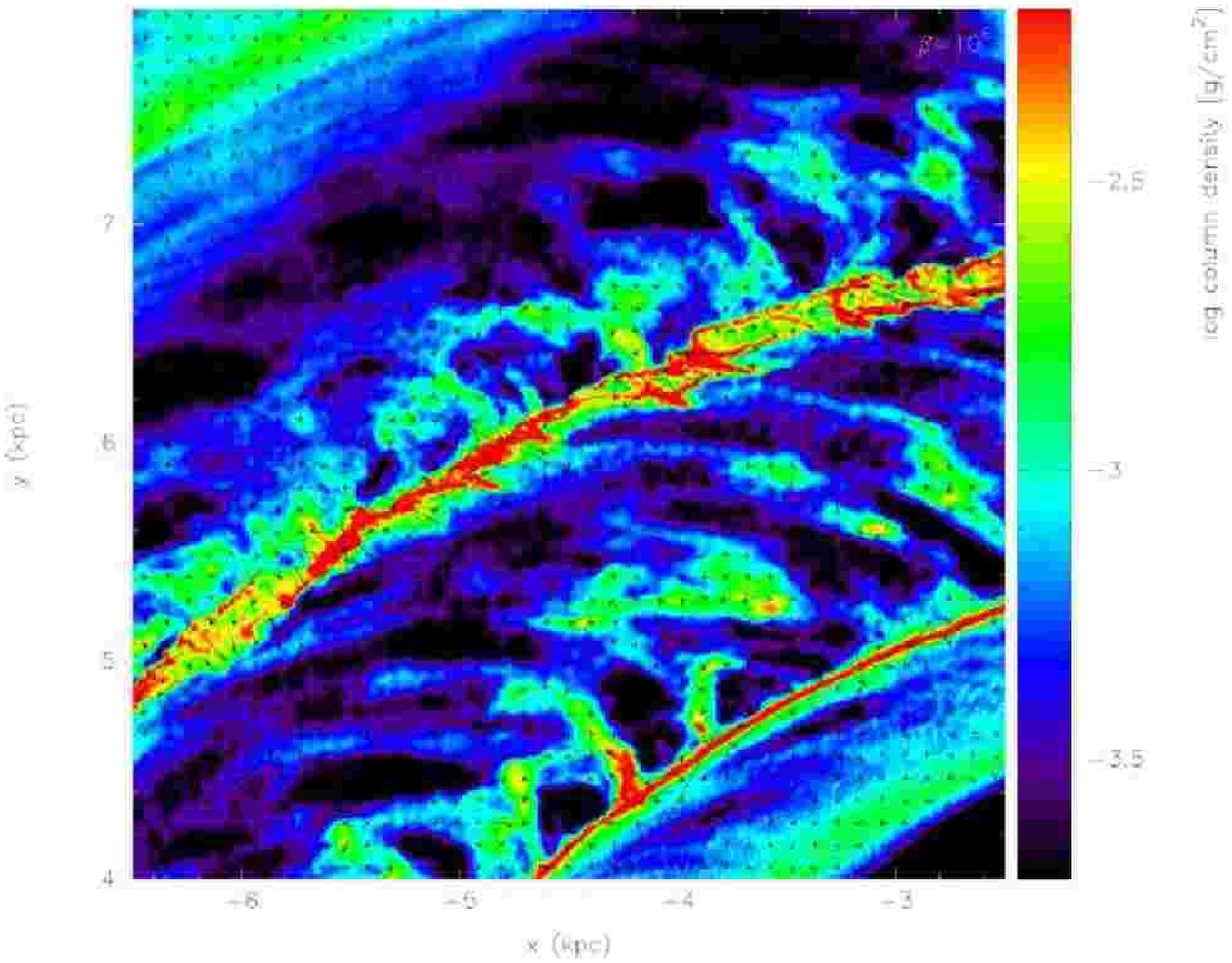}
\includegraphics[scale=0.38,angle=270]{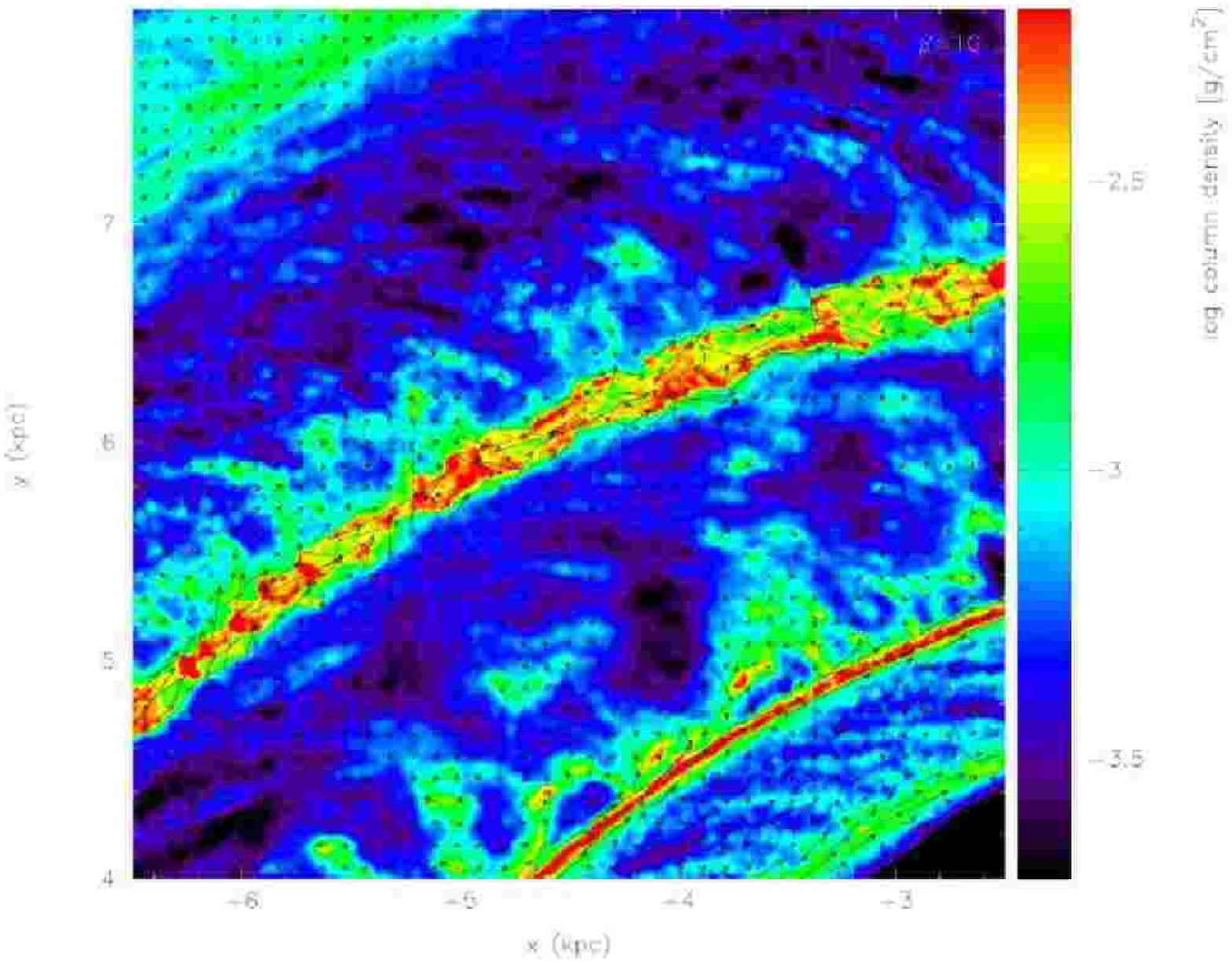}}   
\centerline{
\includegraphics[scale=0.38,angle=270]{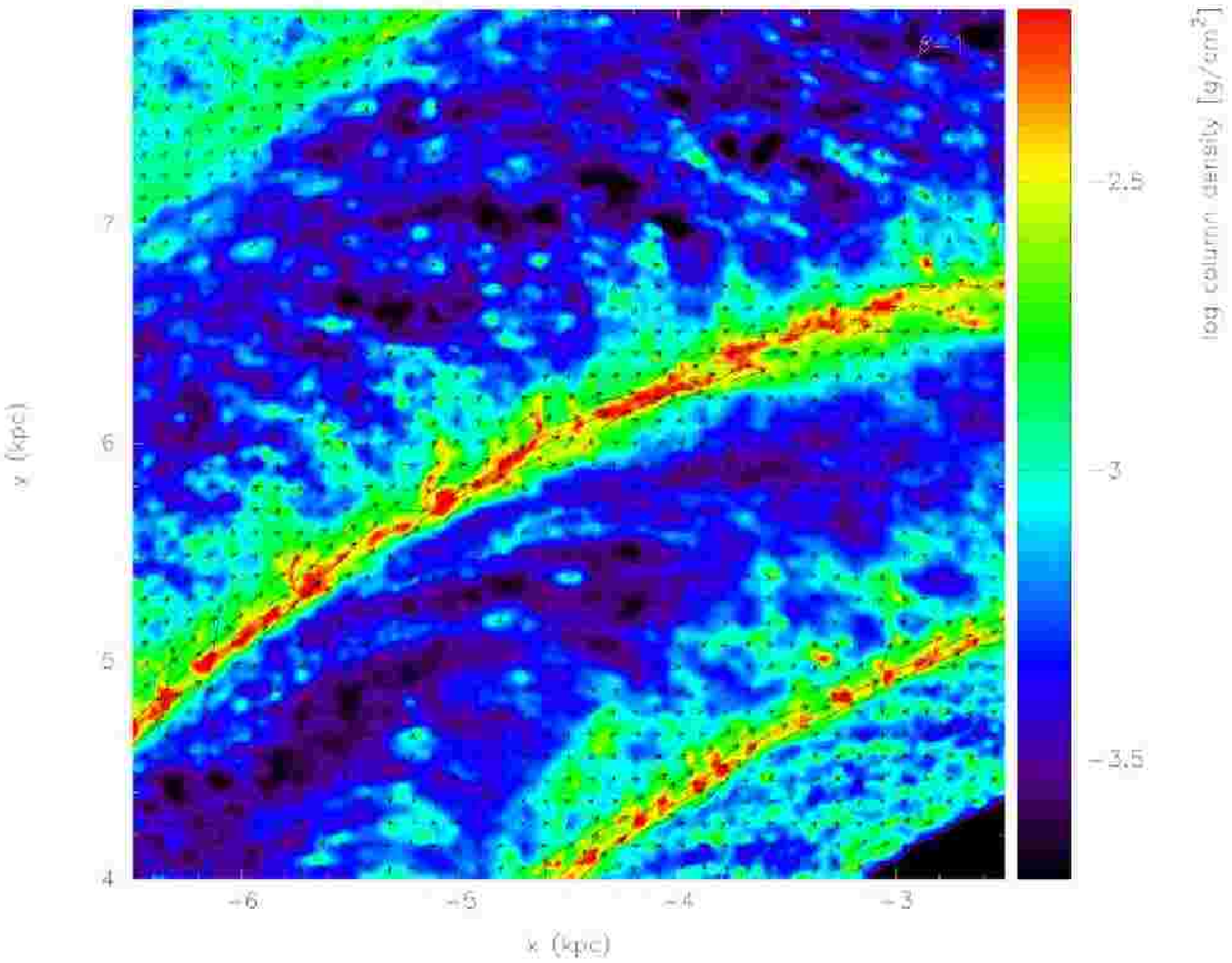}
\includegraphics[scale=0.38,angle=270]{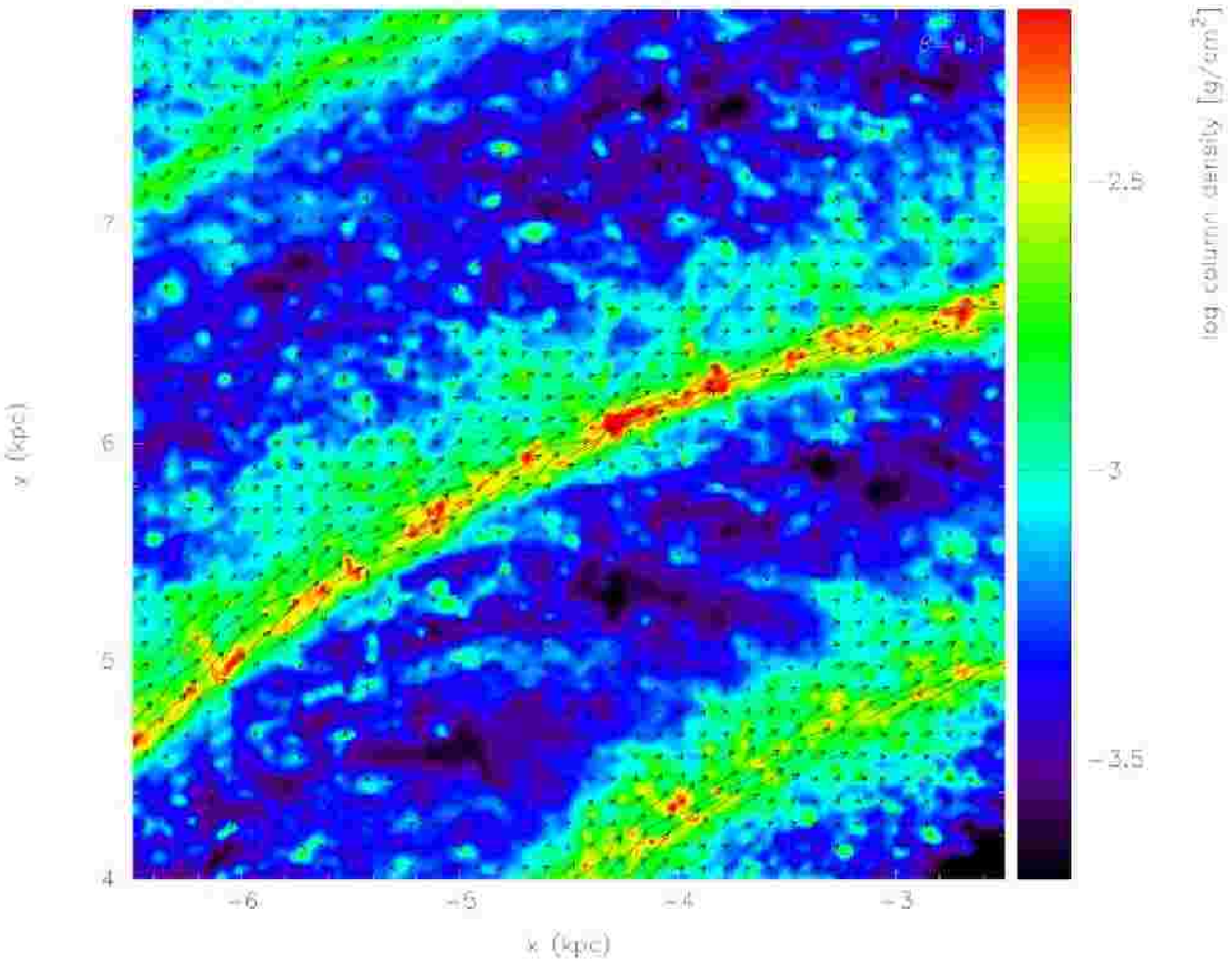}}
\caption{The spiral arm structure of the disc is shown for a 4 kpc x 4 kpc subsection of the disc shown in Figure~1, where the gas is cold.  The ratio of the thermal to magnetic pressure, $\beta$, for each plot is $10^6$ (top left), 10 (top right), 1 (bottom left), 0.1 (bottom right). The inter-arm structure becomes less distinct for higher magnetic field strengths, while for low field strengths, the magnetic field is much more disordered.}
\end{figure*}

\subsubsection{Spiral arm and inter-arm structure}
We show the column density  for the calculations with 100 K gas in Fig.~1. The panels (left to right, top to bottom) denote increasing magnetic field strength, or equivalently, decreasing $\beta$ (as indicated in each panel). The degree of structure and the density of gas in the spiral arms clearly decrease as the magnetic field increases. For the model with the weakest field, which is essentially hydrodynamic, there is considerable inter-arm structure and the spiral arms themselves show distinct clumps (Fig.~1, top left). The formation of spurs perpendicular to the arms and structure within the spiral arms has been previously described in terms of the orbits of gas through the spiral shock \citep{Dobbs2006,DBP2006}. Orbit crowding leads to gas of different velocities interacting in the shock and agglomerating into clumps.
When $\beta=10$, the substructure in the spiral arms is reduced, but spurs are still evident, and the density in the spiral arms is still relatively high.
There are clumps along the spiral arms and substructure perpendicular to the spiral arms when $\beta=1$ (magnetic pressure $\sim$ gas pressure) but the density of these features is much less, and the inter-arm structure is much less distinct. For the case of a strong magnetic field, with $\beta=0.1$ (Fig.~1, bottom right), there is a strong contrast to the models with a weak field (Fig.~1, top). There is little substructure and the spiral arms are much more continuous and smooth. The spiral arms are also much broader for the higher magnetic field strengths.

Subsections of the disc for these results are shown Fig.~2, which more effectively highlight the inter-arm features and magnetic fields. The density contrast of the inter-arm structures is clearer for the cases where the magnetic field is weaker than the gas pressure (top panels). 
These features are more distinct, and between the two inner arms, attain higher densities. 
For the $\beta=0.1$ and $\beta=1$ runs (stronger field), the structure of gas leaving the arms is much more continuous, hence the spiral arm appears broader.
The reduction in the strength of the shock also means that the inter-arm regions are more dense.
Consequently the density of the inter-arm structure when $\beta$ is 0.1 does not deviate significantly from the mean density in the disc and general clumpiness of the gas. 
By comparison there are relatively empty areas in the inter-arm regions in the essentially hydrodynamic case and when $\beta=10$. 

The results from simulations using a gas temperature of $10^4$ K are shown in Fig~3. The top two panels show the case for $\beta=1$. The shock is much weaker (in fact the gas barely shocks at all) compared to when the gas is cold, and the spiral arms are completely smooth and continuous. The structure for the simulations where $\beta=100$ is not significantly different from non-magnetic runs (see \citealt{Dobbs2006}) in which there is also no substructure, but the shock is much weaker when $\beta=1$ due to the magnetic field.
\begin{figure}
\centerline{
\includegraphics[scale=0.3,angle=270]{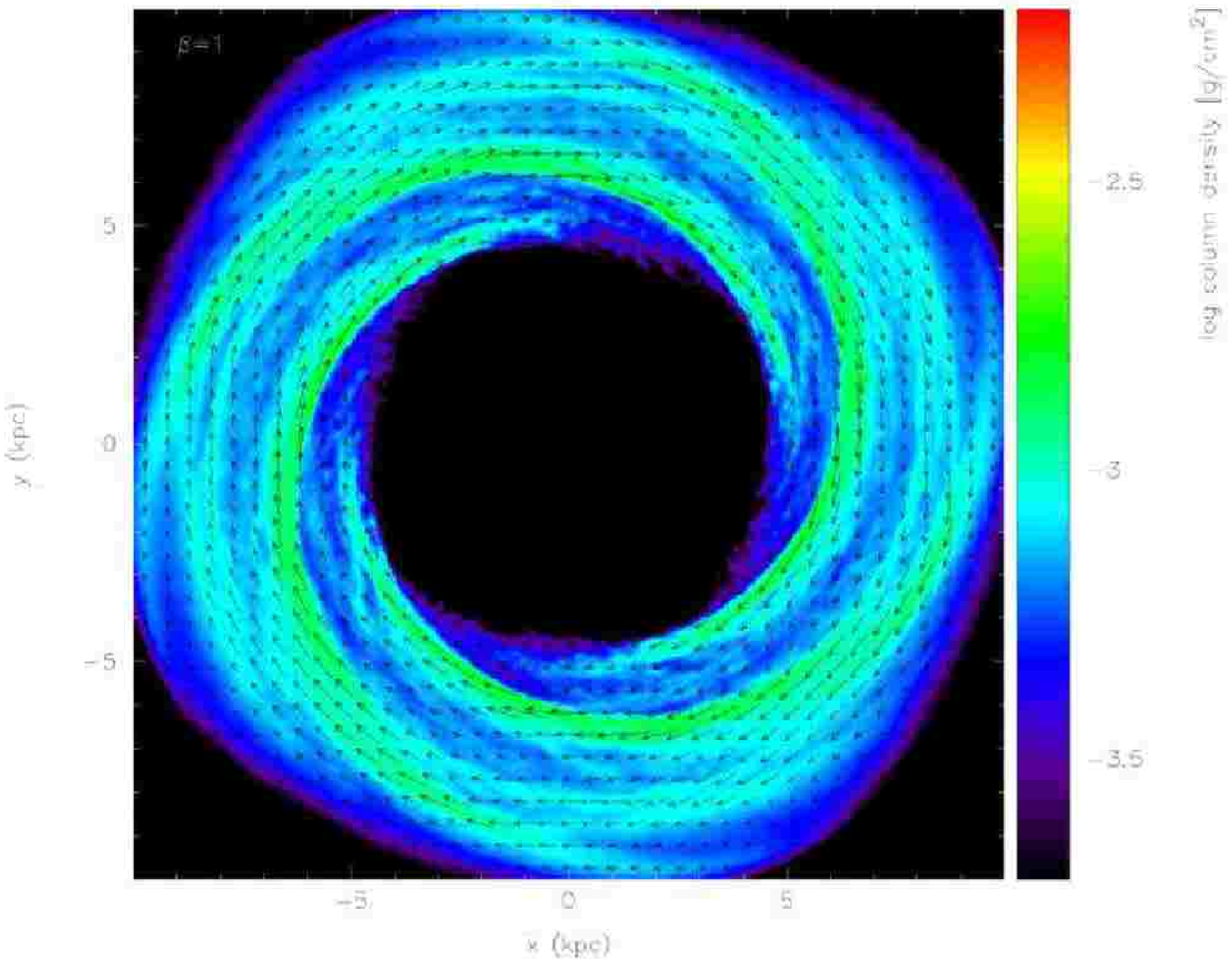}}
\centerline{
\includegraphics[scale=0.3,angle=270]{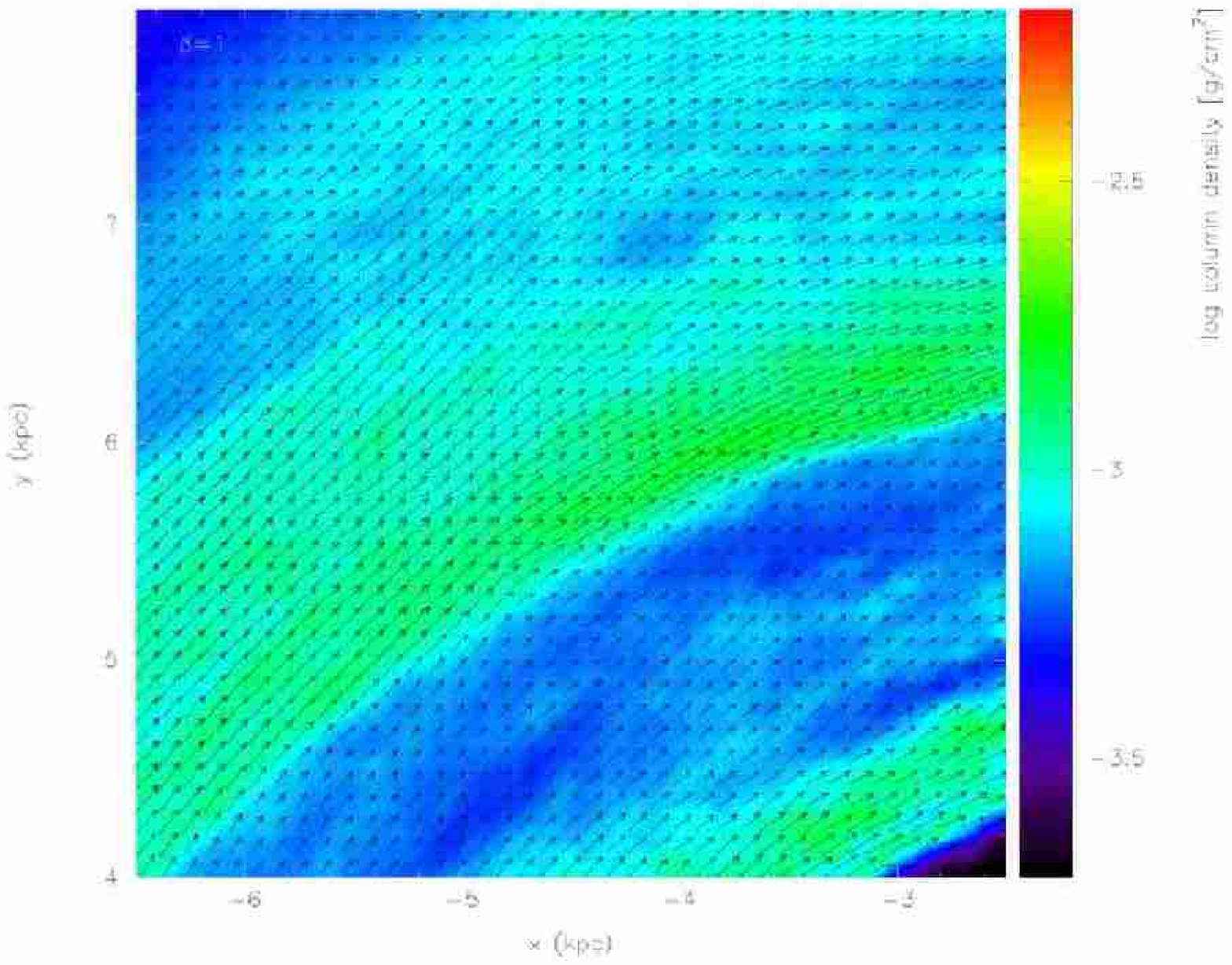}}
\centerline{
\includegraphics[scale=0.3,angle=270]{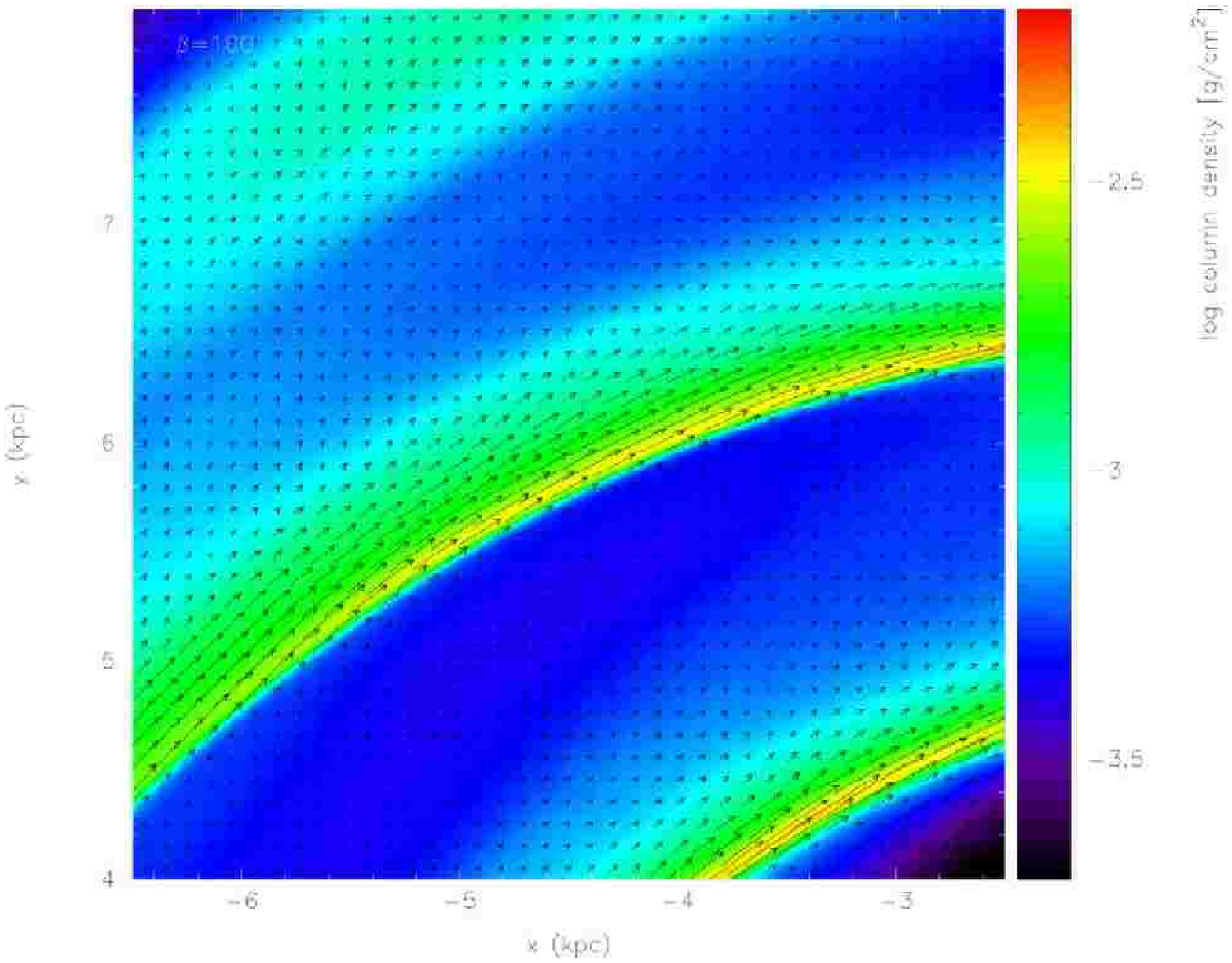}}
\caption{The spiral arm structure of the disc is shown for the single phase simulations with $10^4$ K gas  after 250 Myr. The top panel shows the whole disc when $\beta=1$ (i.e. magnetic and gas pressure equal), whilst the bottom 2 panels show a subsection when $\beta=1$ (middle) and $\beta=100$ (weaker field, bottom). The spiral arms are much smoother for the warm gas, but the spiral shock is very weak when the field is stronger ($\beta=1$). For both field strengths the field is strongly aligned with the gas flow.}
\end{figure}

\subsubsection{Relation of the disc structure to the magnetic and thermal pressure}
The difference in structure between the magnetic and non-magnetic cases is largely attributable to the addition of magnetic pressure (Eqn.~6). When the magnetic field is weak, this pressure is minimal and a strong shock occurs, similar to the hydrodynamic case. If $B$ is large, the magnetic pressure reduces the strength of the shock and smoothes out the gas. This is similar in effect to increasing the thermal pressure, which is also found to reduce the structure in the disc \citep{Dobbs2006}. 

For the case of a weak magnetic field, where $v_A$ is low, the degree of structure depends primarily on the sound speed, as indicated for the non-magnetic case in \citet{Dobbs2006}. 
When magnetic fields become important, the strength of the shock depends on both $c_s$ and $v_A$.
Although we show both warm and cold gas simulations with $\beta=1$, there is more structure in the case where the gas is cold since $c_s$ and consequently $v_A$ are lower. 
We performed a further lower resolution simulation (not shown) with cold gas, taking $\beta \sim 0.01$,  in order to give the same value of $v_A$ as for the model with warm gas and $\beta=1$. In this case magnetic fields dominate thermal pressure much more than expected from observations, but the structure is similar to that shown in Fig~3, top. Thus when magnetic fields are significant, increasing the magnetic pressure by a factor of 100 has an equivalent effect to increasing the thermal pressure by the same degree. 
Assuming that the thermal and magnetic pressures are equal (i.e. $\beta \sim 1$) we find the magnetic field has a strong effect in the simulations with cold gas, but does not remove all the substructure.  
 
\subsubsection{Azimuthal profile and location of shock}
We plot the density against azimuth for the models with cold gas for $\beta=0.1,1$ and $10^6$, as well as the model with warm gas and $\beta=1$, in Fig.~4. To determine the average density we take gas in a ring situated at a radius of 7.5 kpc and width 200 pc which is divided into 200 sections over azimuth. The average density is calculated from the particles in each section.
As expected, the density of the spiral arms increases as $v_A$ decreases and the shock becomes stronger. 
We only show the average densities however - the maximum density is up to an order of magnitude higher than those displayed in Fig.~4.

Also apparent in Fig.~4 is the shift in the location of the shock. The gas shocks further upstream for increasing $v_A$ or magnetic field strength. Similarly the location of the shock occurs further upstream at higher sound speeds \citep{Slyz2003,Gittins2004,Thesis}. Thus increasing the magnetic field strength has a similar effect to increasing the thermal pressure, and the gas shocks earlier. For these simulations we find the gas generally shocks after the potential minimum. This has also been found in other recent results \citep{Gomez2002,Shetty2006}, the former using the same potential as in this paper. We note however that gas appears to shock later for a 4 armed compared to a 2 armed potential \citep{Gittins2004,Thesis}. 
The relative effect of including a magnetic field is to move the shock further upstream. For the case where $\beta=0.1$, and the simulation with warm gas, we find that the shock is roughly coincident with the potential minimum where it would otherwise be downstream of the minimum without the presence of a magnetic field \citep{Thesis}.

\subsection{Evolution of magnetic field for the single phase calculations}
Figures~1-3 also show the galactic magnetic fields in the single phase calculations. 
From Figs.~1 and 3, we see that on galactic scales, the field is strongly aligned with the spiral arms, and compressed in the spiral arms relative to the inter-arm regions. The exception is the case where the magnetic field is very weak ($\beta=10^6$), where the field is  much more random. When there is a stronger field ($\beta \leq 10$), the field appears regular, and follows the large scale gas flow. For the model with warm gas and $\beta=1$, (Fig.~3, top and middle), the very weak shock means there is only a small compression of the field in the spiral arms, and the field is less tightly constrained to the spiral arms.
\begin{figure}
\centerline{
\includegraphics[scale=0.45]{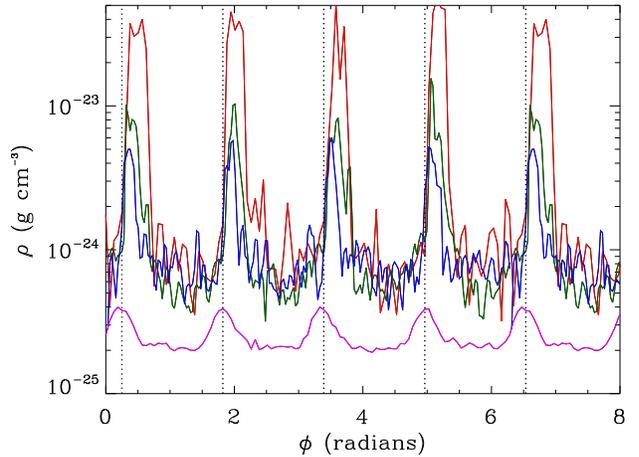}}
\caption{This figure shows the mean density plotted against azimuth for the simulations with cold gas, where $\beta=0.1$ (blue), 1 (green) and $10^6$ (red), and for the simulation with warm gas where $\beta=1$ (magenta). The densities are calculated in the reference frame of the potential after 250 Myr. The azimuthal angle $\phi$ is measured clockwise, in the direction of the flow. The dotted lines indicate the minima of the potential. The density in the spiral arms decreases with increasing field strength, and the shock moves further upstream with respect to the potential minimum.}
\end{figure} 

With a higher value of $\beta=100$, and again a gas temperature of $10^4$ K, we find the morphology of the field is similar, but there is a stronger shock and a greater enhancement of the field in the spiral arms. 
In both simulations of warm gas, the field is very regular in the arm and inter-arm regions at both large and small scales.

On smaller scales, (Fig.~2), we see that the magnetic field is not completely regular for the simulations with cold gas, even with the strongest magnetic field ($\beta=0.1$). We describe how the field disorder is generated in~Section 3.3. In all cases except where $\beta=10^6$, the field is generally aligned along the spiral ams. However in the inter-arm regions, the field is more disordered. For $\beta=0.1$ and 1 (Fig.~2, lower panels), the field still largely follows the gas flow in the spiral arms, but there are significant deviations in the inter-arm regions, particularly when $\beta=1$ and in some places the field is aligned in opposite directions to the large scale gas flow.
As the field strength decreases ($\beta$=10, Fig.~2, top right), the magnetic field becomes more disordered, even along the spiral arms. The field between the spiral arms is much weaker, and shows little coherence.
For the essentially hydrodynamic case, ($\beta=10^6$)  the field appears random everywhere, although considerably enhanced along the spiral arms. At smaller size scales than shown in Fig.~2, the field is actually found to be more confined to the plane of the shock, where the gas is compressed, although the field is aligned in both directions along the shock (c.f. \citealt{Laing1980}). 
If we zoom in to smaller size scales than indicated in Figure~2, the field appears more disordered in the spiral arms for the simulations with cold gas.

\subsubsection{Azimuthal profile}
Fig.~5 shows the average magnetic field strength plotted against azimuth after 200 Myr. The field strength is indicated for the simulations with cold gas where $\beta=0.1$ and $\beta=10$, and the simulation with warm gas and $\beta=1$. We only consider the magnetic field in the plane of the disc here, i.e. $|B|=\sqrt{B_x^2+B_y^2}$. We again determine the average field from a ring of gas at 7.5 kpc as described in the previous section. 
For the simulations with 100 K gas, the field in the spiral arms is around 6 $\mu$G  when $\beta=0.1$, and around 5 $\mu$G when $\beta=10$ (and 4 $\mu$G when $\beta=1$, not shown). The inter-arm magnetic field strengths are $\sim 1 \mu$G when $\beta \leq 1$.
These are average values however - peak magnetic field strengths are $\sim10\mu$G when $\beta=0.1$ and $\beta=10$, and $\sim20\mu$G when $\beta=1$ for the cold gas.
Although for the simulations with higher $\beta$, the magnetic field is weaker, the relative compression of the field in the spiral arms is greater.
For the simulation  with warm gas, with a very weak shock, $B$ increases by a factor of 2. This compares with factors of 5 and 8 when $\beta=1$ and 10 respectively, i.e. for progressively weaker fields. For the model with the very weak magnetic field, the mean magnetic field increases by up to 2 orders of magnitude in the spiral arms, although the mean strength in the spiral arms is only 0.1 $\mu$G. 
The shift in the positions of the peaks in Fig.~5 again reflects the position of the shock with respect to the potential minimum. 

The relative amplitude of the magnetic field in the spiral arms is therefore related to the strength of the shock, as determined by $c_s$ and $v_A$. We have already shown (Fig.~4, Section~3.1) that the density increase in the spiral arms is much greater when the field is weaker. Thus the increase in the magnetic field strength in the spiral arms will also be greatest when the magnetic field is weak.
When considering solutions to an oblique shock \citep{Priest1982}, similar to the passage of gas through a spiral arm here, the ratio of the post and pre-shock magnetic field tends to the ratio of the post and pre-shock densities as $v_A/c_s \rightarrow 0$. 

\subsubsection{$B$ vs $z$}
Recent observations from cosmic ray electrons suggest that magnetic fields exceeding  2 $\mu$G are found up to at least 300 pc above the mid-plane of our Galaxy, and are therefore stronger than expected from the thermal pressure at these latitudes \citep{Cox2005}. We find that the magnetic pressure largely follows the thermal pressure above and below the plane of the disc. Thus the magnetic field does not extend to such large scale heights, but we do not include supernovae or feedback in our simulations.  
\begin{figure}
\centerline{
\includegraphics[scale=0.45]{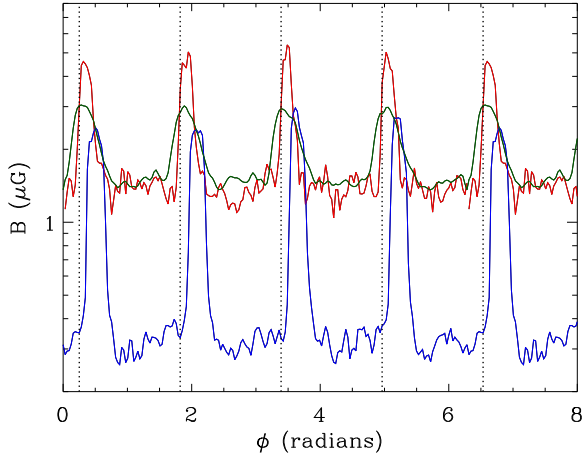}}
\caption{The average magnetic field vs azimuth for the simulations with cold gas where $\beta$ is 0.1  (red), 10 (blue) and with warm gas where $\beta=1$ (green). The corresponding time is 200 Myr.}
\end{figure}

\subsubsection{$B$ vs $\rho$}
We further checked for a relation between $B$ and $\rho$, a property observed in the magnetic fields of  galaxies and molecular clouds. For the gas in the spiral arms, we find $B \sim \rho^{0.7 \pm 0.1}$ when $\beta=1$ and the gas is cold (the exponent is slightly higher, $\sim$ 0.75-0.8 for stronger shocks and lower, $\sim 0.65$ for weaker shocks). This is comparable to the relation expected for compression of a random magnetic field, in which case, by equating the magnetic and thermal pressure, $B \sim \rho^{2/3}$. For our simulations with cold gas, the magnetic field entering the shock has a significant random component assuming the gas has traversed at least one spiral shock. An alternative interpretation is that 
the relation is due to compression in the shock, with $B$ scaling as $\rho$ when $B$ is perpendicular to the shock whilst $B$ is independent of $\rho$ when parallel to the shock.  \citet{Gomez2004} also find $B \propto \rho^{0.7}$, which they relate to isotropic compression.  Observations suggest $B \propto \rho^{0.46}$ for higher density gas, but for gas at lower densities, the exponent is much smaller \citep{Vallee1995}. 

\subsubsection{The ratio of thermal to magnetic pressure, $\beta$}
Although we take initial conditions of a uniform magnetic field and uniform density, at later times in our simulations, $\beta$ takes a range of values about the initial value. However, much of the gas has a value of $\beta$ lower than the average for the disc, particularly in the spiral arms. For example, where the initial $\beta=1$, after 250 Myr, most of the gas in the spiral arms has $\beta \sim 0.1$, so the magnetic field dominates the gas pressure. In the inter-arm regions, the gas and magnetic pressure are more equal (so $\beta \sim 1$). Similarly for the case where the case where the initial $\beta$ is 10, much of the gas in the spiral arms has $\beta$ between 0.1 and 1.
Since $\beta \propto \rho/B^2$ and we typically find $B$ varies as $\rho^{0.7}$, $\beta$ tends to be lower for the dense gas in the spiral arms. Thus the initial values of $\beta$ indicated in Table~1 (and likewise Table~2) only give a very approximate indication of $\beta$ for the gas.

\subsection{The generation of a random magnetic field by spiral shocks}
Galactic magnetic fields are known to have both regular and random components (e.g. \citealt{Beck1996}). Processes which generate turbulence in the ISM, i.e. stellar feedback, gravitational instabilities and MRI are expected to increase the random component of the field, but we focus on the effect of spiral shocks in this paper.

We quantify the degree of order in the magnetic field by resolving the field into $x$ and $y$ components in the plane of the disc. The ordered and random components of $B$ are calculated according to
\begin{align}
& B_{ord} =  \sqrt{<B_x>^2+<B_y>^2}, \nonumber\\
& B_{rand} = \sqrt{<(B_x-<B_x>)^2+(B_y-<B_y>)^2>}
\end{align}
so that the total field is
\begin{align}
& B_{tot} = \sqrt{B_{ord}^2+B_{rand}^2}.
\end{align}
However unlike the previous section, we volume average the magnetic field by integrating over the $z$ direction of the disc. This is similar to Faraday rotation measurements and observations from synchrotron emission. We take 200 points situated evenly in azimuth round the disc at a radius of 7.5 kpc, and consider a cube of sides 100 pc centred on each point. The magnetic field in the $B_x$ and $B_y$ directions are found by integrating over the $z$ dimension at 100 x 100 points, and the averages taken over each set of $100^2$ points. 

Overall, we find that the magnetic field becomes more disordered as the strength of the shock increases, due to the increased velocity dispersion of gas in the shock.
For a simulation where there is no spiral potential, the field remains toroidal and of uniform strength for multiple orbits regardless of $c_s$ or $v_A$. In our calculations with warm gas (where the shock is comparatively weak) the random component of the field is very small. For example, when $\beta=1$ the random component is of order 1 per cent of the total field, whilst with $\beta=100$ the random component is larger, but less than 10\% of the total field.
However simulations with cold gas show a significant random component of the magnetic field, which increases as the magnetic field strength decreases, and dominates the ordered component for 
$\beta \geq 10$.
\begin{figure}
\centerline{
\includegraphics[scale=0.45]{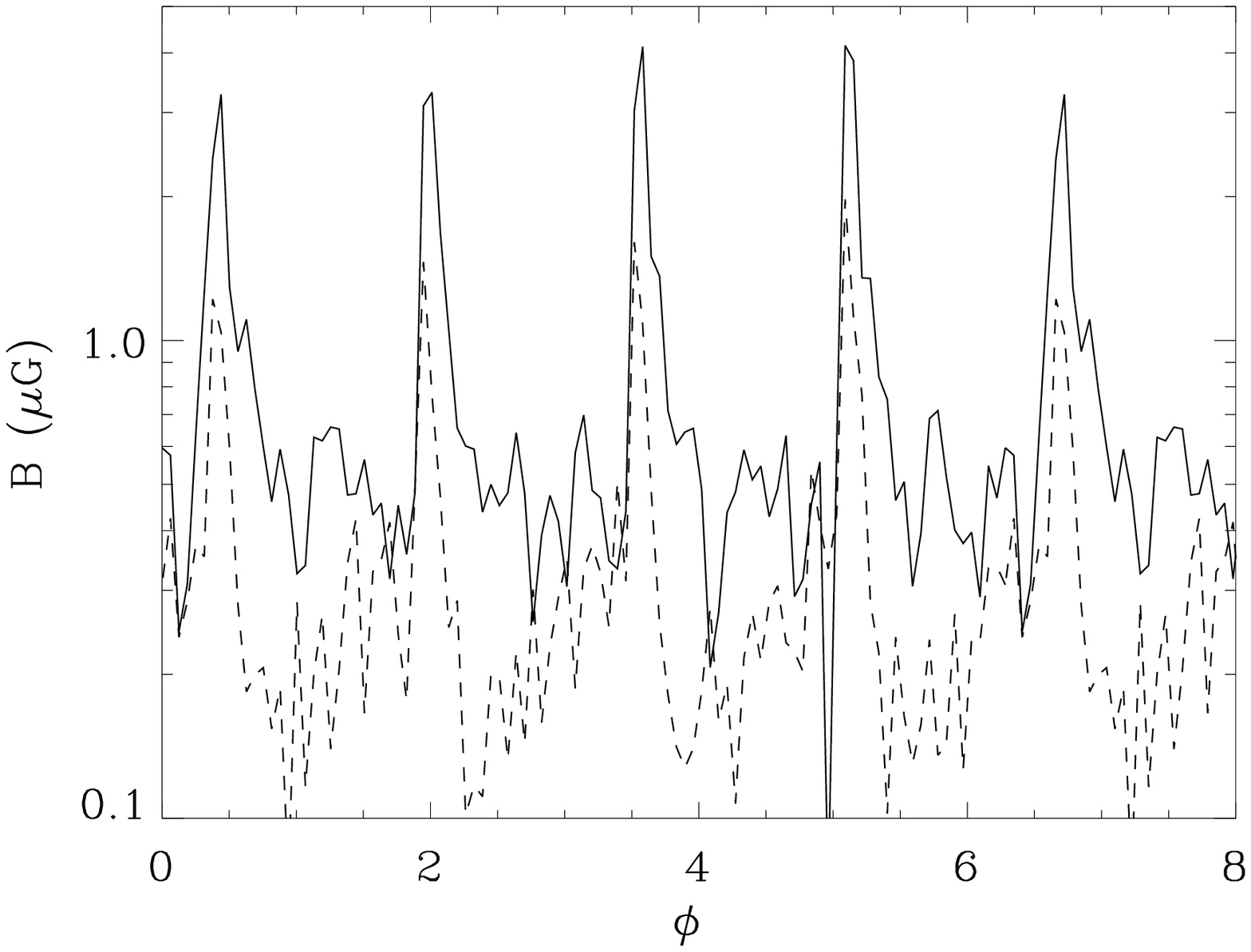}}
\vspace{8pt} 
\centerline{
\includegraphics[scale=0.45]{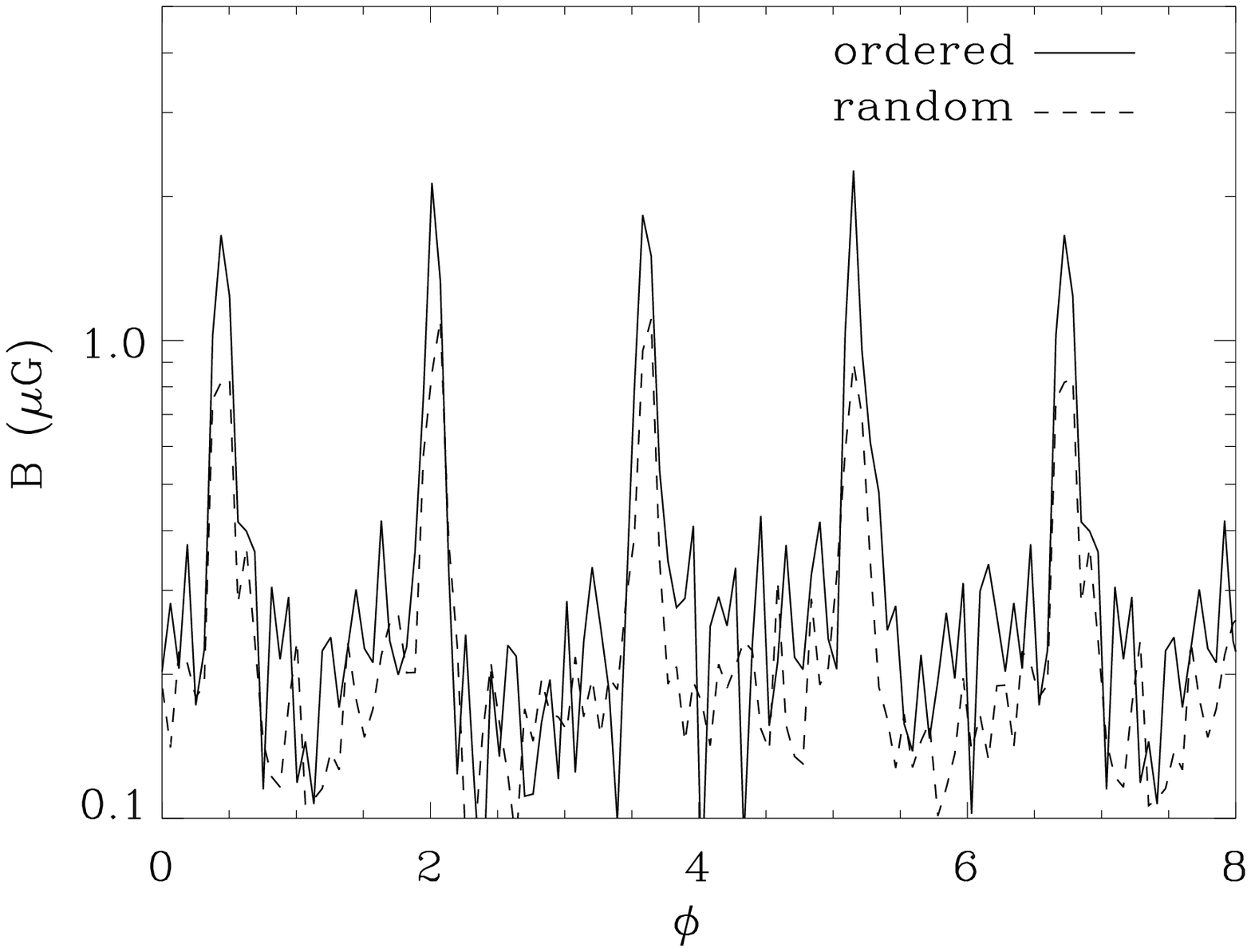}}
\caption{The volume averaged random and ordered components of the magnetic field are plotted against azimuth for the models with cold gas for the strongest magnetic field ($\beta=0.1$, top) and when the magnetic and thermal pressure are equal ($\beta=1$, lower). The corresponding time is 200 Myr. The magnetic field is comparatively more ordered when the field is stronger. }
\end{figure}
\begin{figure}
\centerline{
\includegraphics[scale=0.4]{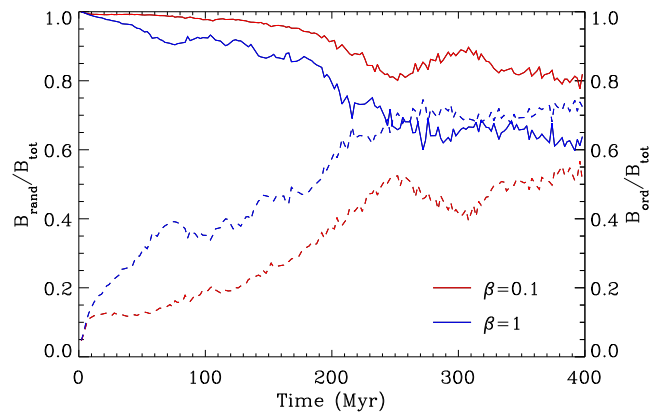}}
\caption{The change in the ordered (solid) and random (dashed) components of the magnetic field, as a fraction of the total field are shown versus time for the models with cold gas for strong ($\beta=$ 0.1) and moderate ($\beta=1$) magnetic fields. The magnetic field becomes more disordered up to around 250 Myr as the gas passes through successive spiral shocks.}
\end{figure}
\begin{figure}
\centerline{
\includegraphics[scale=0.45]{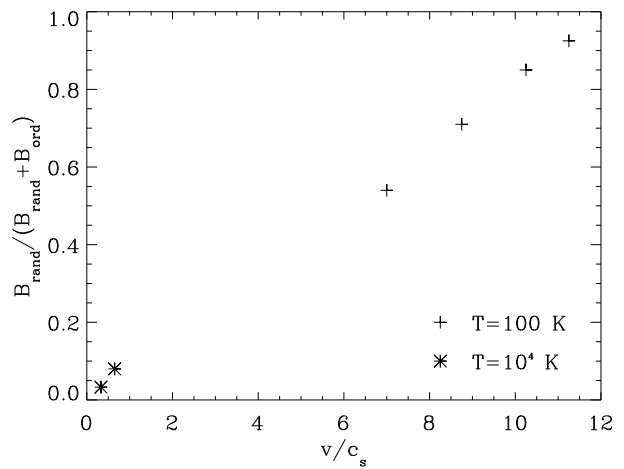}} 
\caption{The random component of the magnetic field versus the velocity dispersion induced in gas in the shock after 250 Myr, for single phase cold (+) and warm (*) simulations. The random component generated is much larger when the velocity dispersion in the spiral arms is high. The points correspond to models A-F in Table~1.} 
\end{figure}

Fig.~6 shows the random and ordered components of the field for simulations with cold gas, with $\beta=0.1$ and 1 after 200 Myr. 
For the strongest field ($\beta=0.1$, top), the ordered magnetic field exceeds the random field in both the spiral arms and inter-arm regions by a factor of 2-3. When $\beta=1$ (producing a stronger shock) the random and ordered components are similar in the inter-arm regions, whilst the ordered component is 1-2 times higher than the random component in the spiral arms. 

The degree of order in the magnetic field also changes with time, as shown in Fig.~7. 
The ordered component decreases as the gas passes through multiple spiral shocks.  The magnetic field becomes disordered in the centre of the disc first, as the evolution of this material is comparatively faster and the gas experiences a shock earlier. 

Note that the ratios of random to ordered fields depend somewhat on the averaging procedure.
For example when using a mass weighted average for $\beta=1$, the measured random component is a factor of 3-4 larger and thereby exceeds the ordered component. 
The random component of the field is reduced in the volume weighted average, since the field is most disordered in the high density regions at low scale heights. The mean field is also reduced by a factor of 1/2  compared to Fig.~5 when using the volume weighted average.

Recent simulations have shown that a velocity dispersion can be induced in the ISM when clumpy gas passes through a spiral shock \citep{Bonnell2006,Dobbs2007a}. In \citet{DBP2006}, we describe the dynamics of the shock as gas interacts and gains and loses angular momentum. For the MHD simulations presented here, the shocks not only induce a velocity dispersion, but the velocities in the gas then lead to a disordered magnetic field. 
For a low resistivity plasma (i.e. ideal MHD) the magnetic field is expected to follow the flow of the gas. Therefore the field will be become tangled where different velocities occur in a localised region of gas. We observe this in simulations where cold gas which has passed through the shock has a much more random field.  
We thus propose that spiral shocks generate and amplify a random component in the magnetic field.

Fig.~8 shows the random component of the field plotted against the velocity dispersion induced in the shock. The dispersion was calculated for the velocity component in the plane of the disc, over subsections of a 200 pc width ring as was used to determine azimuthal profiles of the average density and magnetic field. The velocity dispersions indicated in Fig.~8 are the average from the peak values in each spiral arm. For the simulations with warm gas, the gas entering the shock is fairly uniform, and a supersonic velocity dispersion is not induced in the shock (see \citealt{Dobbs2007a}). By contrast, the velocity dispersion in the simulations with cold gas is highly supersonic, and consequently the random component of the magnetic field is much greater. As the magnetic field increases for a given temperature (i.e. 100 K here), the magnetic pressure acts similarly to thermal pressure in both reducing the strength of the shock, and smoothing out inhomogeneities in the gas. Hence the velocity dispersion in the shock decreases and the field becomes more ordered. 

In some dense regions adjacent to the spiral arms (Fig.~2, lower panels), the field remains ordered, although a velocity dispersion is induced in the gas. This is presumably because there is low shear, so the gas and therefore magnetic field continue to follow the spiral arms. It is only on leaving these regions that the inhomogeneities in the magnetic field are sheared into more circular patterns
\begin{table}
\centering
\begin{tabular}{|c|c|c|c|c|}
 \hline
Model & Gas distribution & $\beta_{cold}$ & $\beta_{warm}$ &  B ($\mu$G)  \\
 \hline
G & 50\% cold, 50\% warm gas &  4 & 400 & 0.1 \\
H & 50\% cold, 50\% warm gas & 0.4 & 40 & 0.3 \\
\hline
\end{tabular}
\caption{Table showing the parameters used for the two-phase calculations, where $\beta_{cold}$ is the initial $\beta$ for the cold gas, and likewise $\beta_{warm}$ for the warm gas.}
\end{table}

\subsection{Two-phase models}
The ISM is known to exhibit a multi-phase medium, so we now turn to models which include both warm and cold gas. Table~2 summarises the parameters for the two-phase models. 
In this paper  we do not include any heating or cooling between the two phases. A resolution study (see Appendix A) shows that the morphology of interarm structure depends on resolution (spurs become more discontinuous and lose their integrity at higher resolution), but the typical spur spacing is independent of resolution.
\begin{figure*}
\centerline{
\includegraphics[scale=0.38,angle=270]{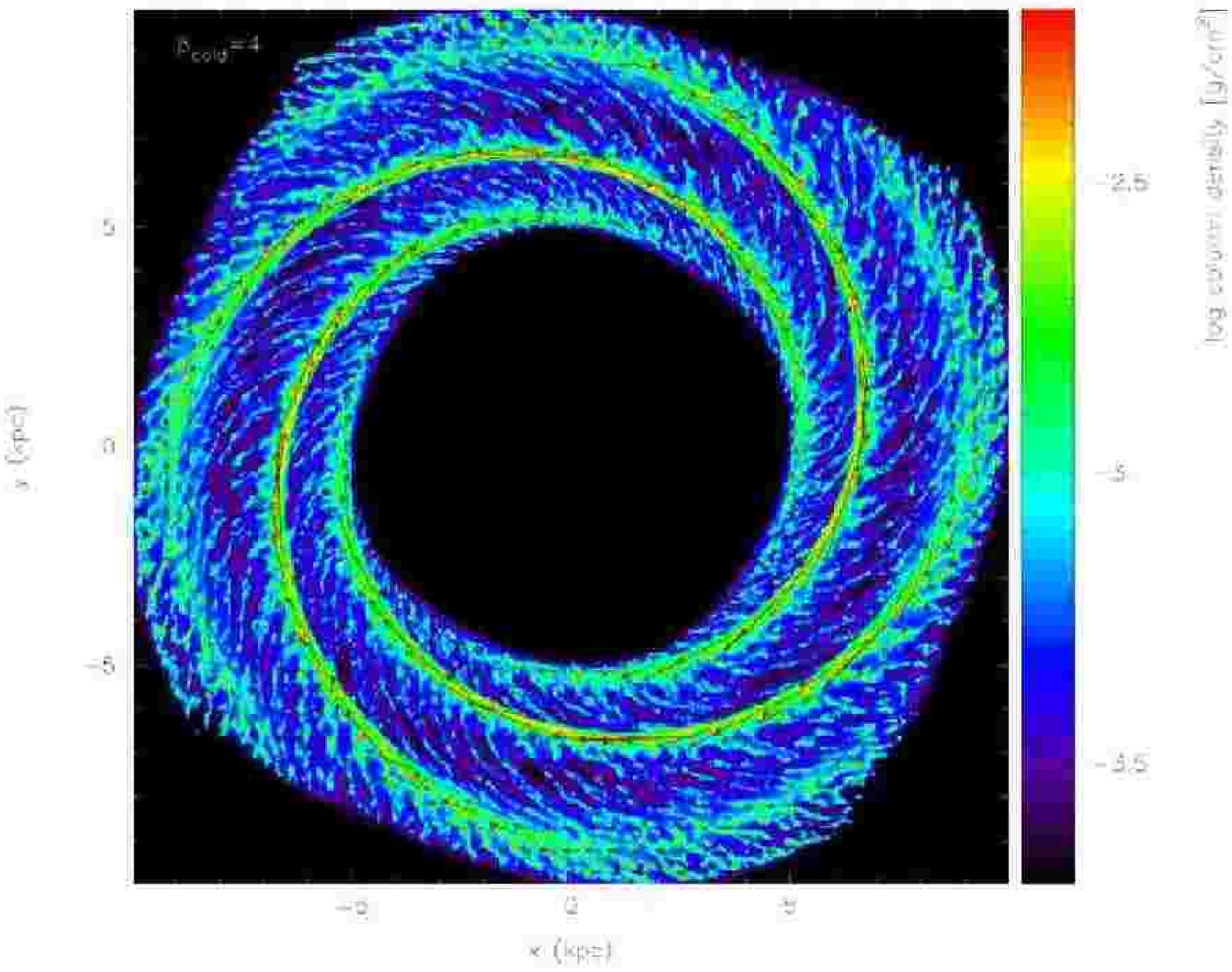} 
\includegraphics[scale=0.38,angle=270]{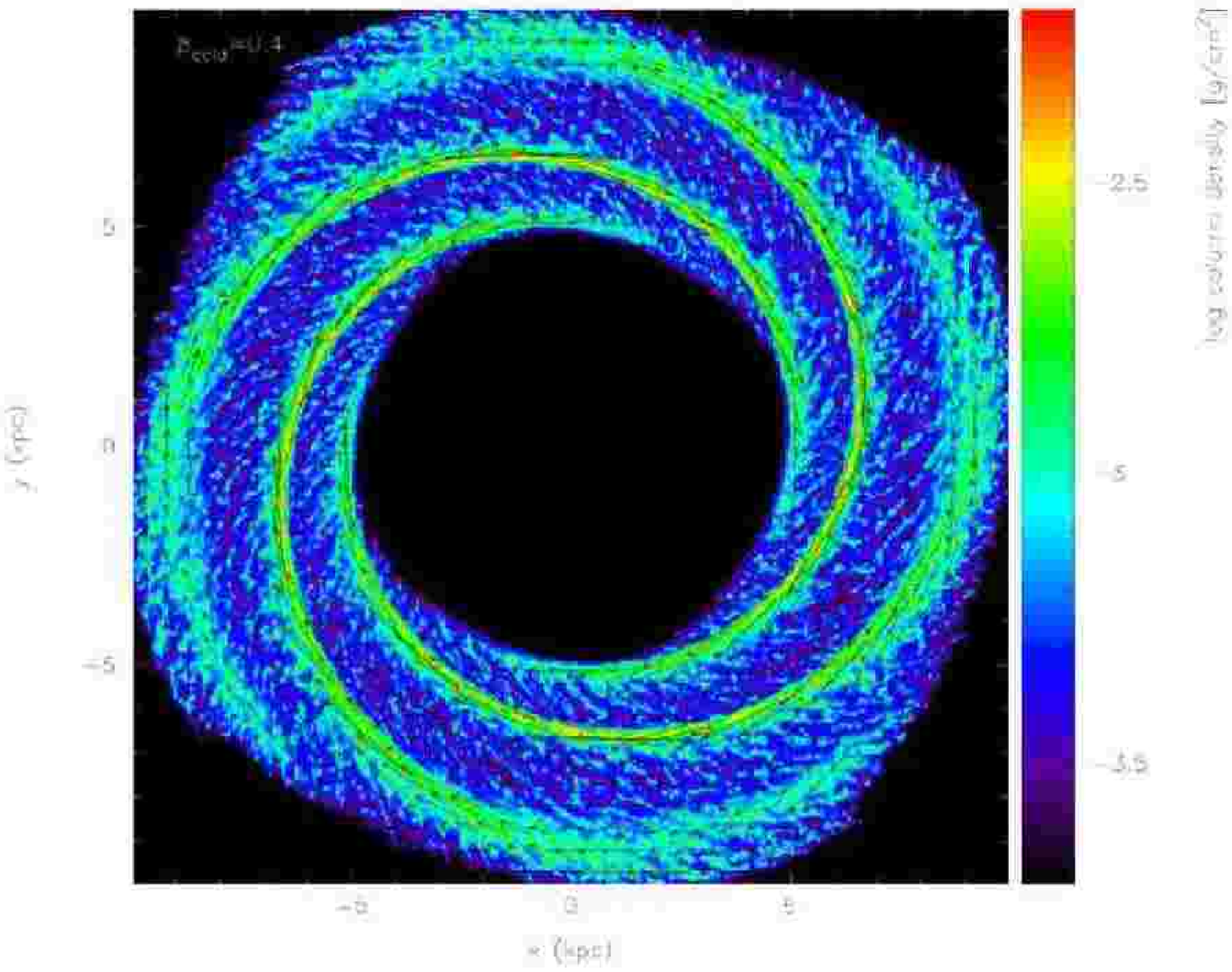}}   
\centerline{
\includegraphics[scale=0.38,angle=270]{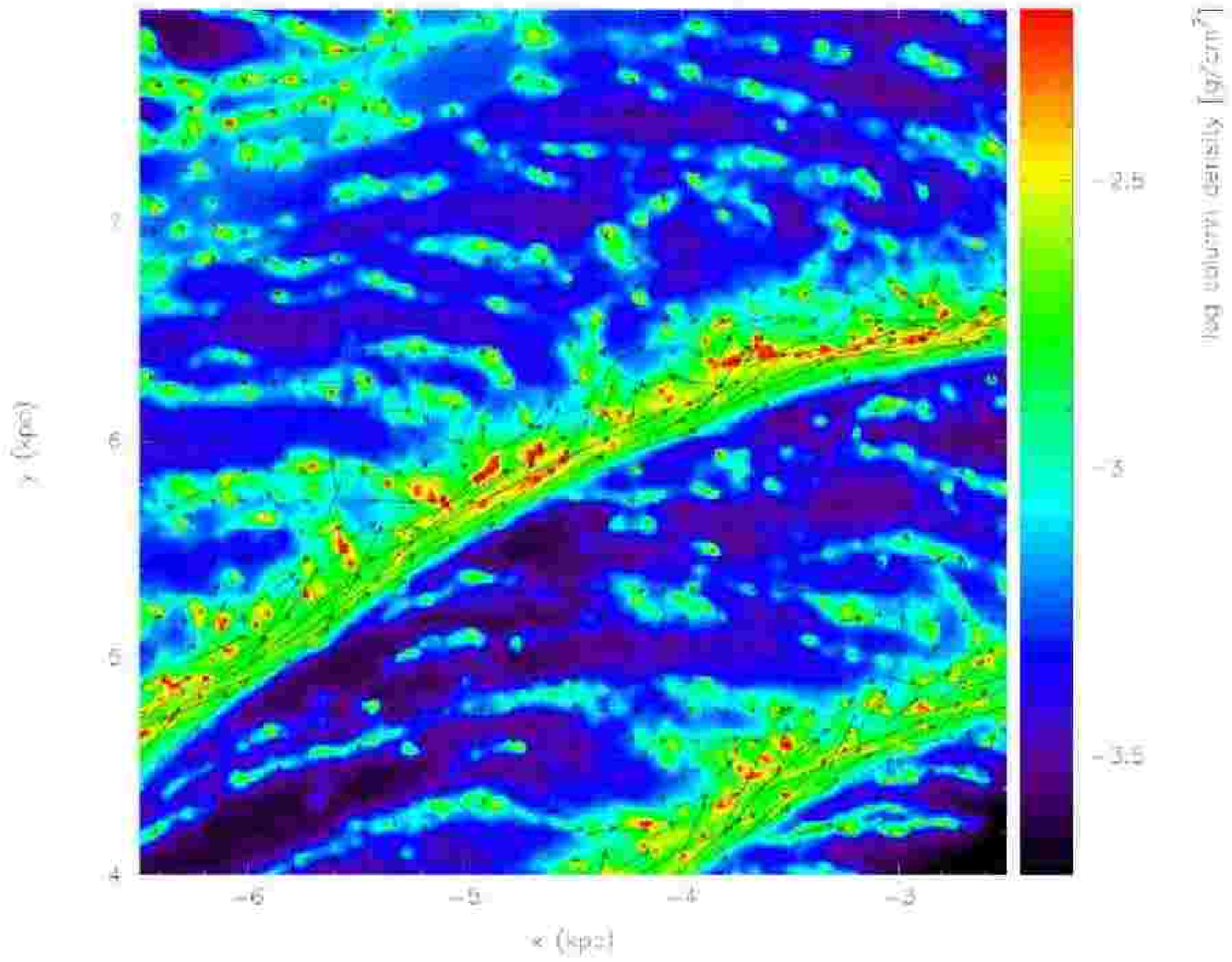}
\includegraphics[scale=0.38,angle=270]{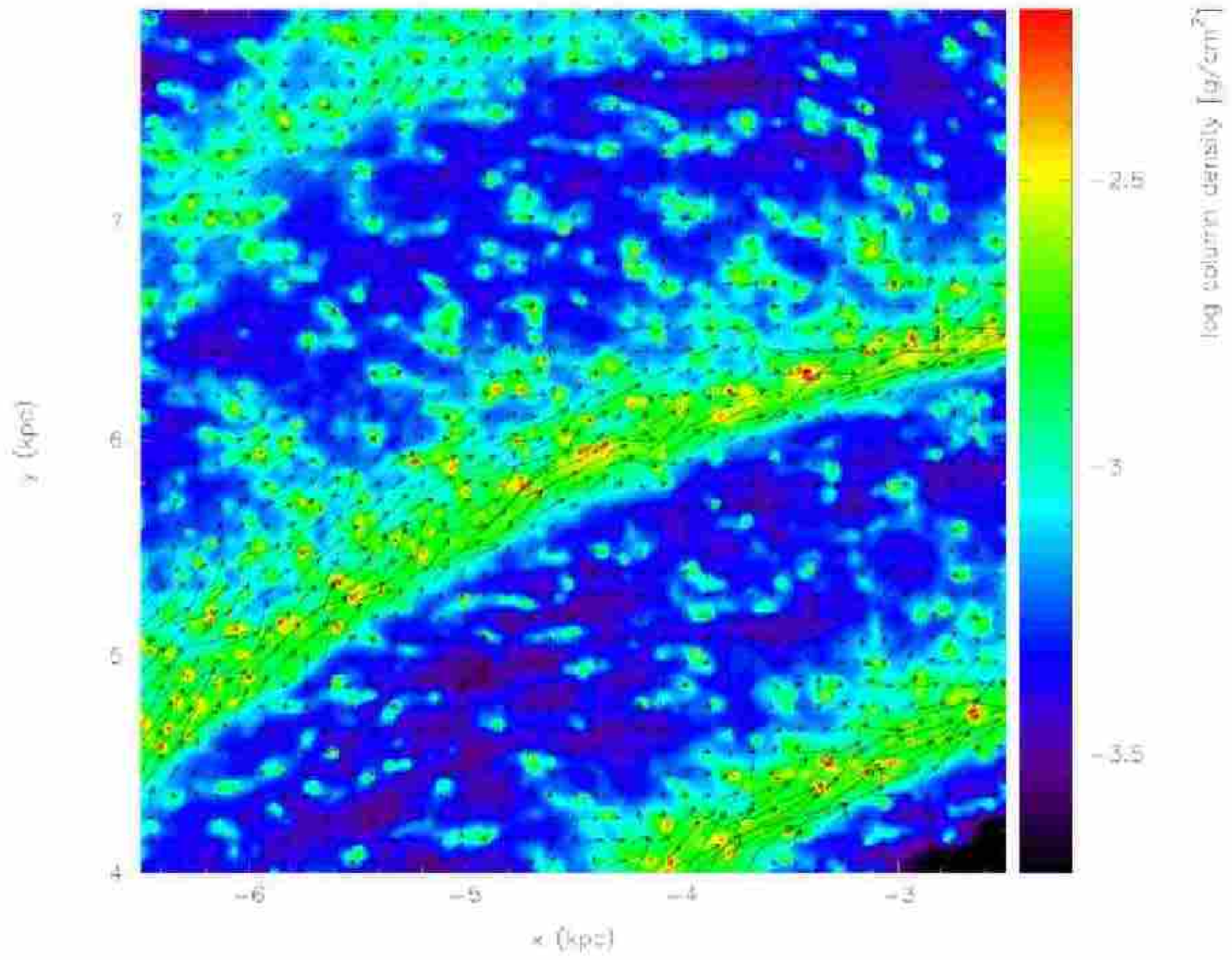}}
\caption{The column density is shown for the two-phase simulations after 250 Myr, for the whole disc (top) and a 4 x 4 kpc subsection (bottom). The left hand panels show the case where $\beta_{cold}=4$  and the right hand panels where $\beta_{cold}=0.4$.
Both the cold and warm phases are shown in the plots, but we show them separately for the case where $\beta_{cold}=4$ in Fig.~12.
There is more structure in the cold gas when the magnetic field is weaker ($\beta_{cold}=4$). The vectors show the magnetic field smoothed over a particular grid size. There is more detailed structure on smaller scales, particularly in the spiral arms which are better resolved.}
\end{figure*}

\subsubsection{Spiral arm and inter-arm structure}
The results for the two phase calculations calculations are presented in Fig~9 with $\beta_{cold}=4$ (left) and $\beta_{cold}=0.4$ (right). The top and bottom panels display the whole disc and a 4 x 4 kpc subsection respectively, the latter coinciding with the sections shown in Fig.~2. In both models the cold gas forms clumps, whilst the warm phase provides a background diffuse medium, similar to previous hydrodynamic simulations \citep{Dobbs2007}. 
The densities of the spiral arms and inter-arm clumps are reduced for the stronger fields (lower $\beta$), whilst for weaker fields (higher $\beta$), the clumps are more coherent and form longer, more spur-like structures. From the left hand panels in Fig.~9, we see that the inter-arm gas is similar to the `rungs of a ladder' between the spiral arms, with dense clumps situated along them. This structure is persistent with time, as demonstrated in Fig.~10, which shows the disc in the $\beta_{cold}=4$ run after 600 Myr.
The structure in this case is much more similar to M51, the archetypal galaxy for comparing inter-arm features, which has long thin spurs dotted with star forming regions.

Whilst the magnetic pressure tends to weaken the spiral shock and smooth out structure, the presence of a warm phase counteracts this to some extent. For the case where $\beta_{cold}=4$, the inter-arm clumps are more dense than the corresponding time in the single phase model where $\beta=10$ (Fig.~2, top right). 
We have previously noted in hydrodynamical simulations that the warm component provides a pressure which confines the cold gas to higher densities \citep{Dobbs2007}. Overall we find from the simulations in this paper, and lower resolution simulations, that if $\beta_{cold} \lesssim 0.1$, the cold and warm phases do not separate (since even in single phase simulations the magnetic pressure smoothes out any structure in the cold gas), but tend to remain much more randomly distributed. With $\beta_{cold} \gtrsim 1$, more coherent features develop in the cold gas, but for intermediate values only smaller clumps tend to form.

Thus in the results presented in this paper, the cold and warm gas are essentially separate, i.e. the cold gas is situated in clumps embedded in the warm phase. This is because the increase in density of the cold gas due to the spiral shock is much larger than that of the warm phase. Orbit crowding in the spiral shock then leads to the agglomeration of these clumps into larger structures, though this process is reduced by magnetic pressure, and is more evident when self gravity is included (Dobbs et. al., in preperation). This occurs regardless of whether the simulations are single or two phase. The densities of the cold gas are higher when a warm phase is present (particularly in the inter-arm regions), but the structure of the cold gas is similar (e.g.  Fig.~9 top and Fig.~2 bottom left).

\subsubsection{Evolution of magnetic field} 
Similarly to the single phase results, the magnetic field appears regular on galactic scales, but more disordered in the inter-arm regions on smaller scales (Fig.~9). There is again a more random magnetic field in the model with a higher $\beta$ (Fig.~9, left, compared to Fig~9, right). 
In Fig.~11 we plot the cold (top) and warm (lower) components separately for the model with $\beta_{cold}=4$. As expected from the single phase results, the magnetic field of the warm gas is more regular than that of the cold, especially in the inter-arm regions. However the magnetic field of the warm gas is clearly more random than in entirely single phase simulations of warm gas (compare to Fig.~3).  

We show the volume averaged ordered and disordered components of the magnetic field for the two-phase simulation with $\beta_{cold}=4$ explicitly in Fig.~12. For the warm gas, the disordered and ordered components of the field are of comparable strengths. This is in contrast to the single phase simulations of warm gas, where the disordered component was much smaller than the total field. Thus the magnetic field in the warm gas is modified by the distortions of the field in the cold gas and the velocity dispersion induced in the cold phase.
%The field in the cold gas in Fig.~11 appears more irregular compared to the single phase simulations (e.g. where $\beta=10$, Fig.~2, top right) since the warm gas confines the cold to higher densities, but the increase in the magnetic field strength is only related to the strength of the shock. 
\begin{figure}
\centerline{
\includegraphics[scale=0.38,angle=270]{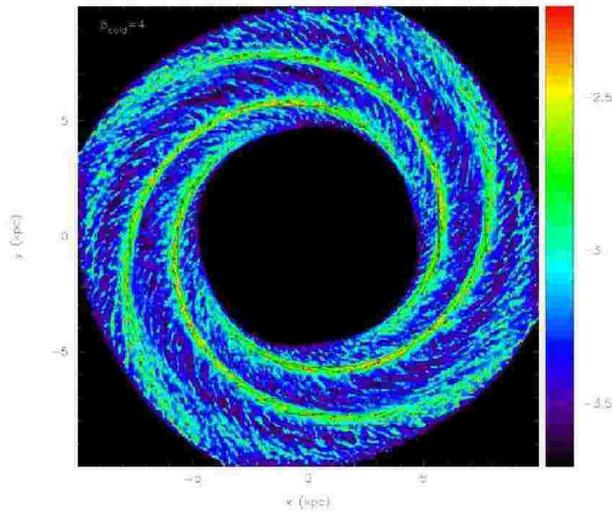}}   
\caption{The column densities and magnetic field for the two-phase simulation with $\beta_{cold}=4$ are shown after 600 Myr.}
\end{figure}
\begin{figure}
\centerline{
\includegraphics[scale=0.38,angle=270]{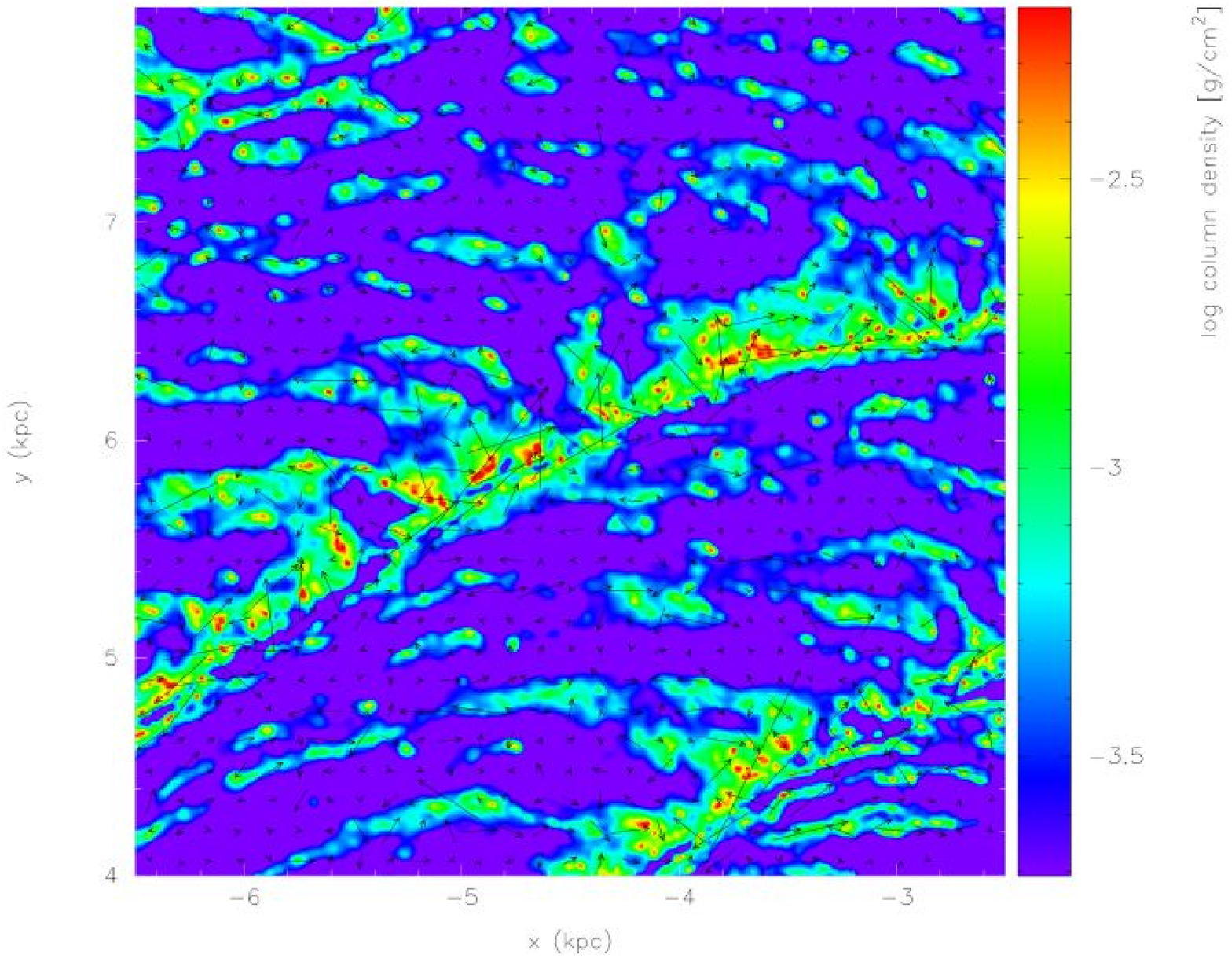}}   
\centerline{
\includegraphics[scale=0.38,angle=270]{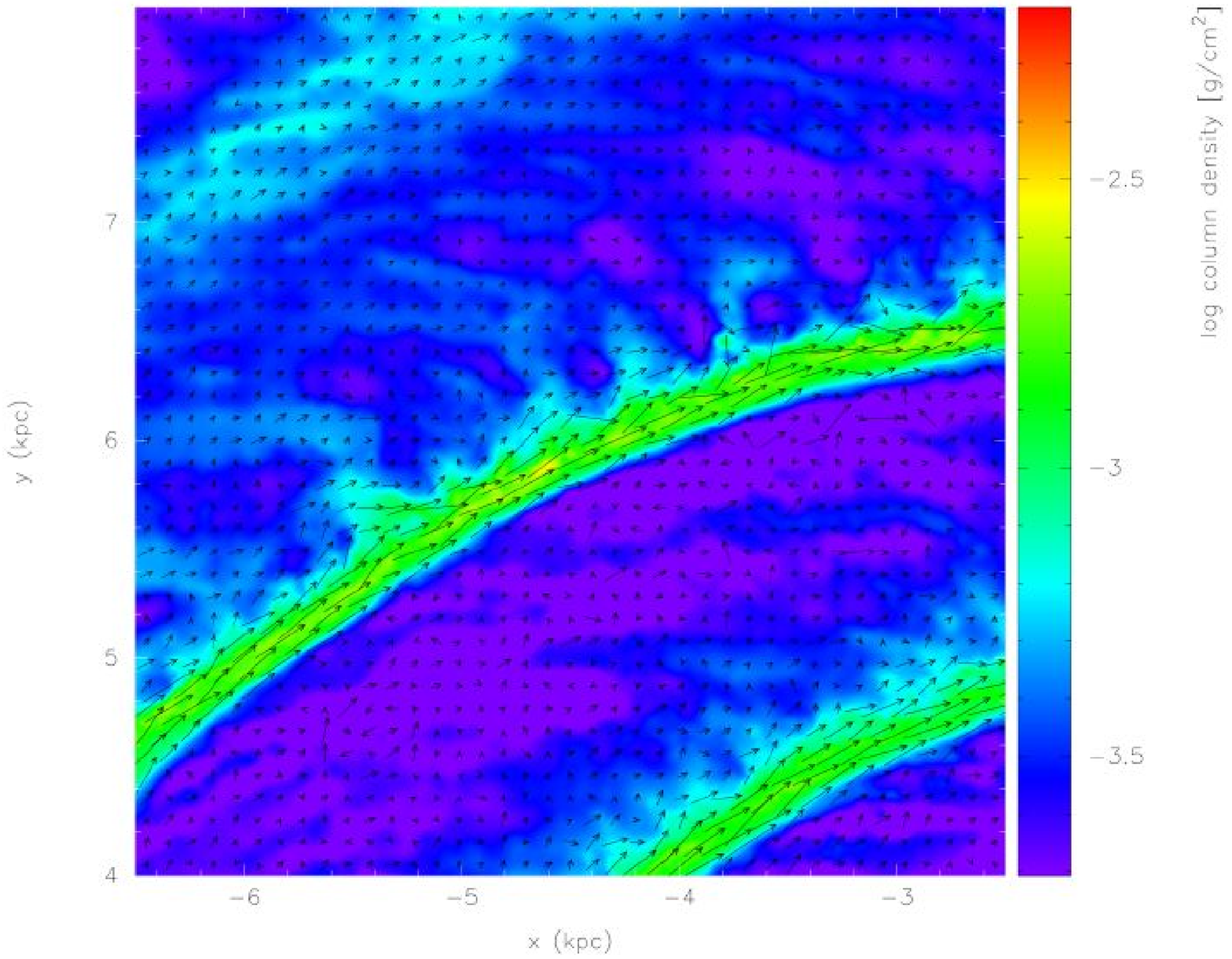}}
\caption{The column densities and magnetic field for the cold (top) and warm (gas) are shown separately where $\beta_{cold}=4$. This 4 x 4 kpc subsection of the disc is the same as the bottom right panel of Fig.~9. The magnetic field is much more disordered in the cold gas compared to the warm component.}
\end{figure}
\begin{figure}
\centerline{
\includegraphics[scale=0.45]{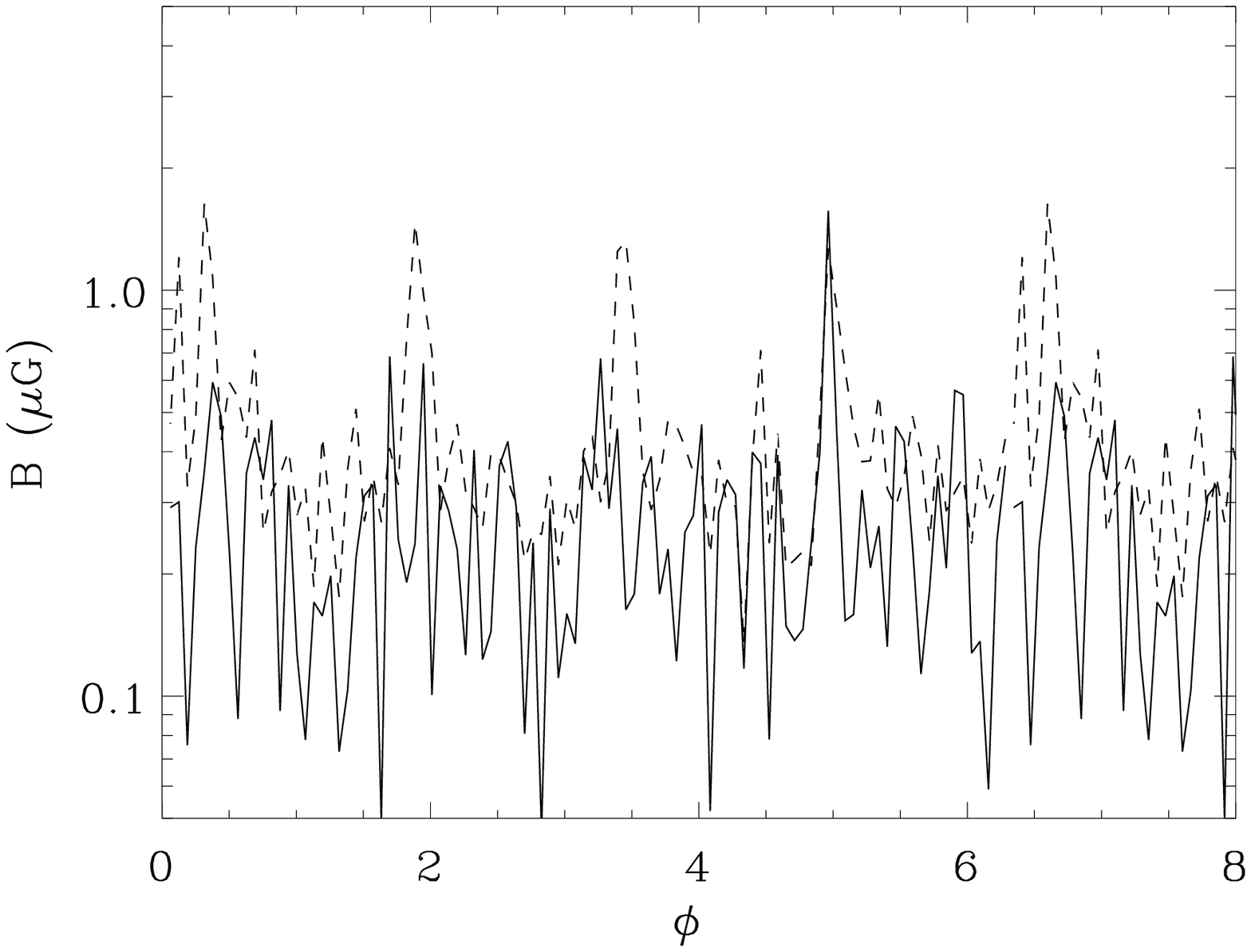}}  
\vspace{8pt} 
\centerline{
\includegraphics[scale=0.45]{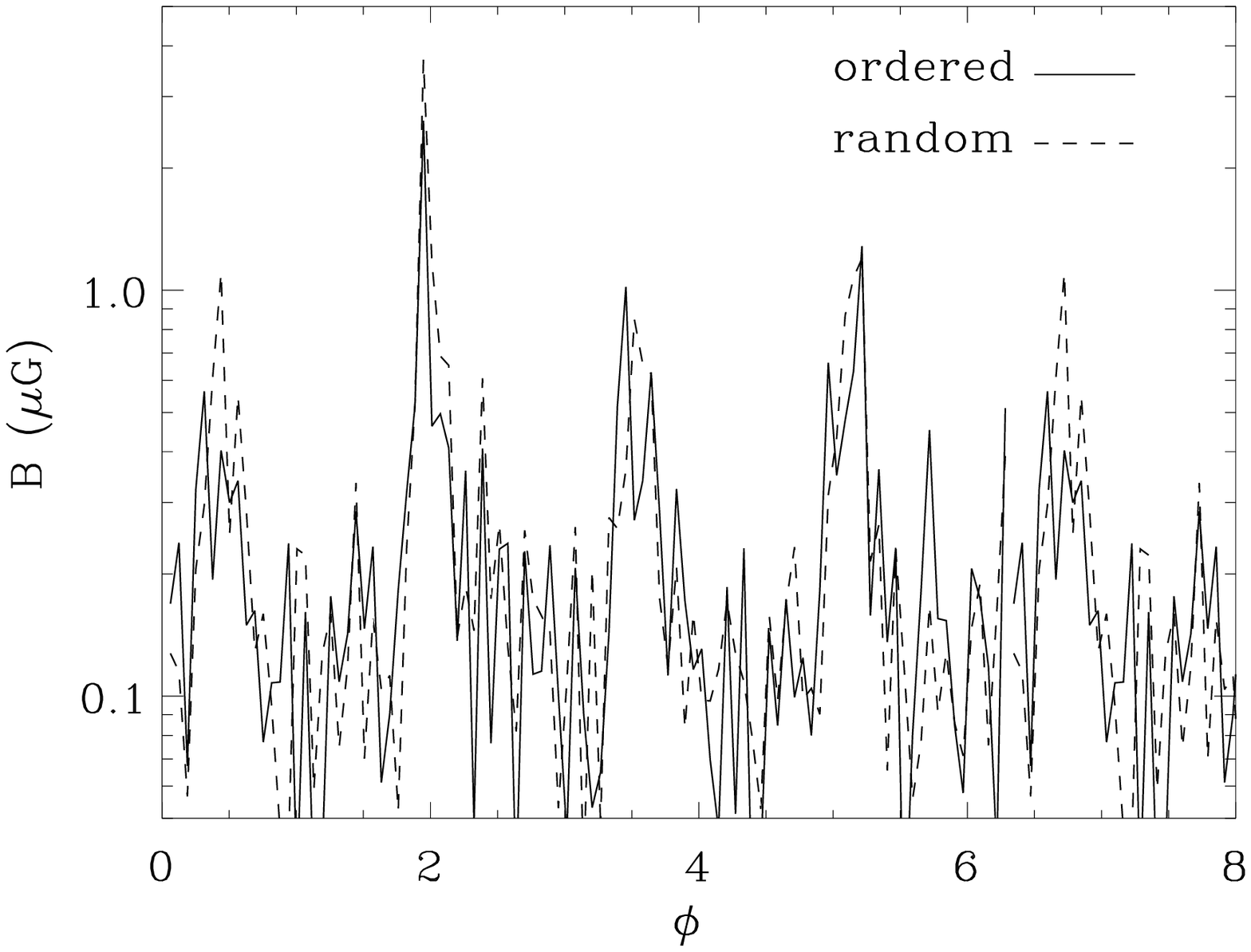}}
\caption{The volume averaged ordered and disordered components of the magnetic field are plotted for the warm and cold gas in the two-phase simulation with $\beta_{cold}=4$ for the cold gas. The top panel shows the magnetic field for the cold gas, whilst the lower panel shows the magnetic field for the warm gas.}
\end{figure}
 
A caveat to these results is that since we took initial conditions consisting of a uniform density disc of uniform magnetic field strength, $\beta$ for the warm gas is 100 times higher than that of the cold (an alternative would be to start with an initially clumpy distribution). As mentioned in Section~2.1, $\beta$ is expected to be similar for the cold and warm components of the ISM. However, a smaller value of $\beta$ for the warm gas is unlikely to change the structure of the disc significantly in the two-phase results since this primarily depends on the cold gas, and the warm phase merely acts as an extra pressure. Although the magnetic field may be more ordered with a lower $\beta$, we find the field is much more disordered when $\beta$ for the warm gas is 40 in the two-phase results compared to $\beta=100$ in the single phase results. Thus it seems likely that a highly irregular field in the cold gas will have some effect on the field in the warm gas even if $\beta$ is lower for the warm gas.

\subsection{Comparison of magnetic field with observations}
\subsubsection{Warn Neutral/Ionized Medium}
The principal technique for measuring magnetic fields in the warm ISM is from radio measurements of diffuse synchrotron emission. Synchrotron emission arises from the interaction of cosmic rays with warm gas and has enabled mapping of magnetic fields in several external galaxies such as M51 and M81 \citep{Beck2007}. An estimate of the distribution of cosmic ray electrons is also required in order to determine the magnetic field strength. For $\lambda=6$~cm, Faraday rotation of the emission is negligible, and radio polarisation provides the strength and orientation of the regular component of the field. In conjunction with Faraday rotation measurements at several wavelengths, the direction of the field can be measured. 

Observations from synchrotron emission are generally associated with the warm component of the ISM. The total 
(random + ordered) magnetic field is typically $\sim 6\mu$G, although magnetic field strengths are higher in the spiral arms of grand design galaxies, with strengths of 10's of $\mu$G \citep{Beck1996}. 
Our simulated field strengths for the warm gas tend to be lower by a factor of 3 to 4, although the maximum field strengths are much more comparable to the synchrotron measurements. The mean strengths are closer to Faraday rotation measurements of pulsars in the Galaxy, which indicate lower field strengths (e.g. $3 \mu$ G, \citep*{Han2006}). However the magnetic field scales with density. In providing values for the magnetic field, we took a surface density comparable to the solar radius, where the magnetic field is relatively weak \citep{Han2006}.
The random and ordered components of the field are thought to be of similar magnitude. This is the case for the two phase simulations, whilst the random component is minimal in simulations with just warm gas.

\subsubsection{Synthetic observations}
In order to make a meaningful comparison of our results with observations of magnetic fields in spiral galaxies, we have computed a synchrotron polarisation map of one of our simulations. A detailed description of how the polarised emission is calculated is described in Appendix B. 

The synthetic polarisation map is shown in Fig.~13 for the two phase simulation where $\beta_{cold}=4$. The polarised emission is enhanced in the spiral arms, whilst the magnetic field is generally aligned with the spiral arms comparable to radio-intensity maps of M51 \citep*{Nein1992,Patrikeev2006,Beck2007}. In our simulations, the spiral shock corresponds to the dust lanes of actual galaxies, generally coincident with trailing edge of the optical spiral arms. However the most notable difference from our simulations is that the polarized intensity in the observations of M51, and similarly M83 \citep{Beck2007} forms much wider spiral arms, with contours of the emissivity spread over a few kpc into the inter-arm regions. By contrast in our simulations, the enhancement of the magnetic field is strongly confined to the regions corresponding to the spiral shock. Some observations of galaxies even find that the magnetic spiral arms do not coincide with the optical spiral arms where star formation is occurring \citep*{BeckNat1996}. This phenomenon is difficult to explain purely by the dynamics of spiral shocks, as is evident from our models. 
\begin{figure}
\centerline{
\includegraphics[scale=0.38,angle=270]{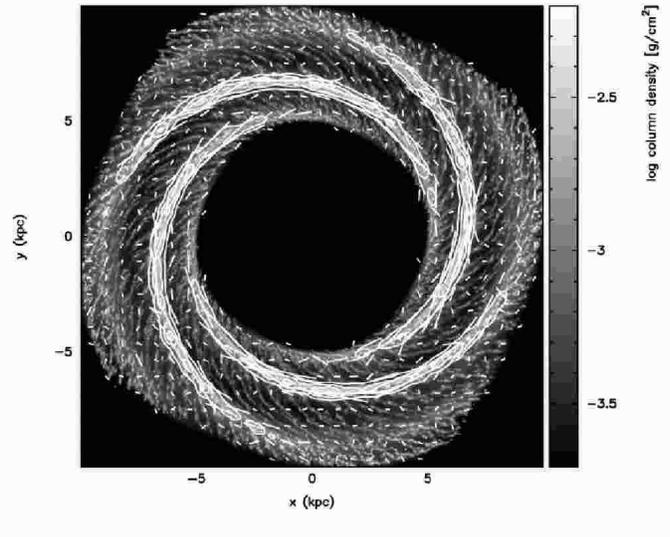}}   
\caption{A synthetic polarisation map for the two phase simulation where $\beta_{cold}=4$. 
The arrows ($B-$vectors) are constructed from the Stokes Q and U parameters and the contours show the total synchrotron intensity (Stokes I parameter) smoothed to a beam size of 0.7 kpc (ie. $h_{beam} =$ 0.35 kpc).}
\end{figure}

\subsubsection{Cold Neutral Medium}
Unfortunately there are no corresponding observations of the magnetic field in cold HI on galactic scales. Zeeman splitting has recently been used to determine magnetic fields in cold HI structures in the Arecibo survey \citep{HT2005}, but this method requires strong magnetic fields. Furthermore the observations do not provide spatial information on the magnetic field, rather a median field strength of $6 \mu$ G. The cold gas in the spiral arms of our simulations has a field strength of around $2 \mu$ G in the two phase simulations and with $\beta=1$. 
The relatively low magnetic field strengths in our simulations suggest that it would be worth setting up simulations where the cold gas is already situated in dense clumps, and establishing whether this structure is maintained for higher field strengths.
We find a comparatively greater increase in the magnetic field strength in the shocks for the cold gas than warm gas, typically by a factor of 5-10.

\citet{LiG2006} calculate order parameters (i.e. the ratio of the ordered to total field) of $\sim0.2-0.4$ in several $\sim100$ pc size GMCs, where the configuration of the magnetic fields in the GMCs is thought to be a consequence of the external magnetic field rather than internal sources. We expect more detailed calculations to show whether GMCs in computational models exhibit a similar degree of order.

\section{Conclusion}
We have performed simulations of galactic discs subject to a spiral potential with a range of field strengths. The main results we have discussed are 1) the reduction in structure across the disc as the magnetic field strength increases, and 2) the possibility of spiral shocks inducing an irregular magnetic field in the ISM. 

As the strength of the magnetic field increases, the strength of the spiral shocks and therefore density of the spiral arms are reduced. This is as expected from the analysis of \citet{Roberts1970}. Consequently the formation of spurs is increasingly suppressed for higher magnetic field strength. The spiral arms themselves are more continuous and clumps along the spiral arms are less dense. 
This supports previous 2D results \citep{Shetty2006,Tanaka2005} which find that both Kelvin-Helmholtz and gravitational  instabilities perpendicular to the magnetic field are reduced by stronger magnetic fields (although we do not relate the structure in our simulations to these instabilities, our overall conclusions agree). We find that the addition of the magnetic field is similar in effect to an increase in thermal pressure, in that both provide a pressure which oppose the formation of structure, and smooth out the gas. 

Nonetheless, we still find significant inter-arm structure with the presence of a magnetic field, unlike the results (those which are non-self gravitating) of \citet{Shetty2006}. This structure is present in the cold component of the ISM, which has generally not been included in previous simulations. Inhomogeneities present in the initial random distribution of gas become amplified by spiral shocks. With warm gas or strong magnetic fields, these inhomogeneities are smoothed out by the pressure. For our calculations with cold gas, we find that magnetic fields only prevent substructure when the ratio of gas to magnetic pressure ($\beta$) is $\lesssim 0.1$. For $\beta \gtrsim 1$, spurs perpendicular to the arm still form in the cold gas. The additional pressure provided by warm gas in the two-phase results increases the longevity of structure in the inter-arm regions, even when the magnetic field dominates for the cold gas.

The main difference between ours and previous results is that we use cold gas, although we also use a weaker magnetic field than \citet{Shetty2006}.
We find structure is present for values of $\beta$ in the cold gas similar to observations, but the volume averaged magnetic field strengths in our two phase models are typically $< 1 \mu$G, lower than observations. This discrepancy arises because we start with an initially uniform distribution of cold gas. To obtain a similar $\beta$ for both the cold and warm gas in our two phase models, we would need the cold gas to be 100 times denser than the warm gas, i.e. this just indicates that structure is \emph{already} present in the cold gas. If we used higher field strengths, we would find, like the results of \citet{Shetty2006}, very little structure emerges. In this case, $\beta$ for the warm gas would be closer to observations, but the magnetic pressure would dominate the thermal pressure by a factor of 100 or so in the cold gas, inconsistent with observations. Thus we either need to distribute the cold gas initially in clumps, or ideally include a much more detailed thermal treatment of the ISM, in order to obtain magnetic field strengths \emph{and} $\beta$ consistent with observations.

We find that whilst the ISM appears highly structured in observations, we would expect a higher degree of structure with relatively weak magnetic fields. For instance, the two phase model where $\beta_{cold}=4$ retains much more structure typical of grand design galaxies compared to the case where $\beta_{cold}=0.4$. Current observations suggest that magnetic pressure exceeds thermal pressure \citep{HT2005}.
We however note that $\beta$ exhibits a range of values in our simulations, and $\beta$ tends to be take smaller values in the spiral arms and dense gas (for a given temperature), which are more likely to correspond with observations.
Again a more complete treatment of the ISM in future work, and further observations of the CNM will allow a better comparison.

In our simulations, the magnetic field is compressed by the spiral shocks, as expected from a straightforward analysis of MHD shocks \citep{Roberts1970,Priest1982}. The relative increase in the magnetic field strength is greater where the shock is stronger.
However the most intriguing result from our simulations is the possibility that spiral shocks generate an irregular magnetic field. This process has not been identified in previous simulations, which we attribute to the fact that they have not included a cold phase. Galaxies are known to contain a random component of the field, but it is usually supposed that this is due to supernovae and/or feedback from stars. 
We therefore postulate that spiral shocks are important in generating disorder in the magnetic field, whilst simultaneously inducing a velocity dispersion and density structure in the gas (\citealt*{Bonnell2006,KKO2006}; \citealt{DBP2006}).
The degree of order in the disc, similar to the presence of arm/inter-arm structure is related to the strength of the shock. When the gas is cold and the magnetic field is weak, the shock is much stronger, and a higher velocity dispersion is induced in the gas. Consequently the magnetic field lines follow the gas and the field becomes tangled. 
In simulations with warm gas, we find the magnetic field is almost entirely regular, although with a stronger shock, the field may become more irregular. However in our two-phase results, the velocities and irregularities in the magnetic field of the cold gas induces comparatively more disorder in the warm gas than the single phase calculations. Although there are no measurements of regular versus disordered components of the magnetic field in cold HI, we would predict from our results that the field in the cold gas will be more disordered than that of warm HI.
\section*{Acknowledgments}
We thank the referee for various helpful comments and suggestions. We also thank Matthew Bate and Jim Pringle for several useful discussions and comments.
This work, conducted as part of the award ÒThe formation of stars and planets: Radiation hydrodynamical and magnetohydrodynamical simulationsÓ made under the European Heads of Research Councils and European Science Foundation EURYI (European Young Investigator) Awards scheme, was supported by funds from the Participating Organisations of EURYI and the EC Sixth Framework Programme. DJP is supported by a PPARC postdoctoral research fellowship.
 
Computations included in this paper were performed using the UK Astrophysical
Fluids Facility (UKAFF). Figures were produced using SPLASH, a visualisation package for SPH that is publicly available from http://www.astro.ex.ac.uk/people/dprice/splash/ \citep{Price2007}.
\appendix
\section{Resolution tests and spacing of spurs}
In order to demonstrate that the length scales of structure in our simulations is not dependent on resolution, we perform a study of the two phase simulations with $\beta_{cold}=4$ at different resolutions. Fig.~A1, shows simulations with 1 (left) and 8 (right) million particles. With higher resolution, the structure tends to become more disjointed, and therefore it is harder to recognise large scale features. However the main spurs which extend across the inter-arm regions tend to be similar at different resolutions. In fact the structure of the cold gas in the two phase simulations is essentially the same as an equivalent simulation with only cold gas. 

The large scale structure and spacing between spurs is determined by the dynamics of the spiral shock, and therefore should not be affected by the numerical resolution. Instead we find that the spacing between spurs is dependent on the strength of the spiral shock, and is thus related to the epicyclic radius, and the thermal and magnetic pressure. Given that the primary purpose of the present paper is to examine the effect of including a magnetic field, we intend to discuss this in an upcoming paper (Dobbs, in prep.) which investigates a range of parameters, including the strength of the potential, the addition of self gravity, and in these calculations, the spacing between spurs is clearer. 

We have also performed calculations where the cold gas is initially distributed in cold clumps, rather than randomly. In this case, the initial structure is somewhat different because the clumps are sheared into thin dense filaments. However as the gas passes through the spiral shocks, this structure tends to be disrupted, and the spacing between spurs is similar to that apparent in Fig.~A1. 
\begin{figure*}
\centerline{
\includegraphics[scale=0.4]{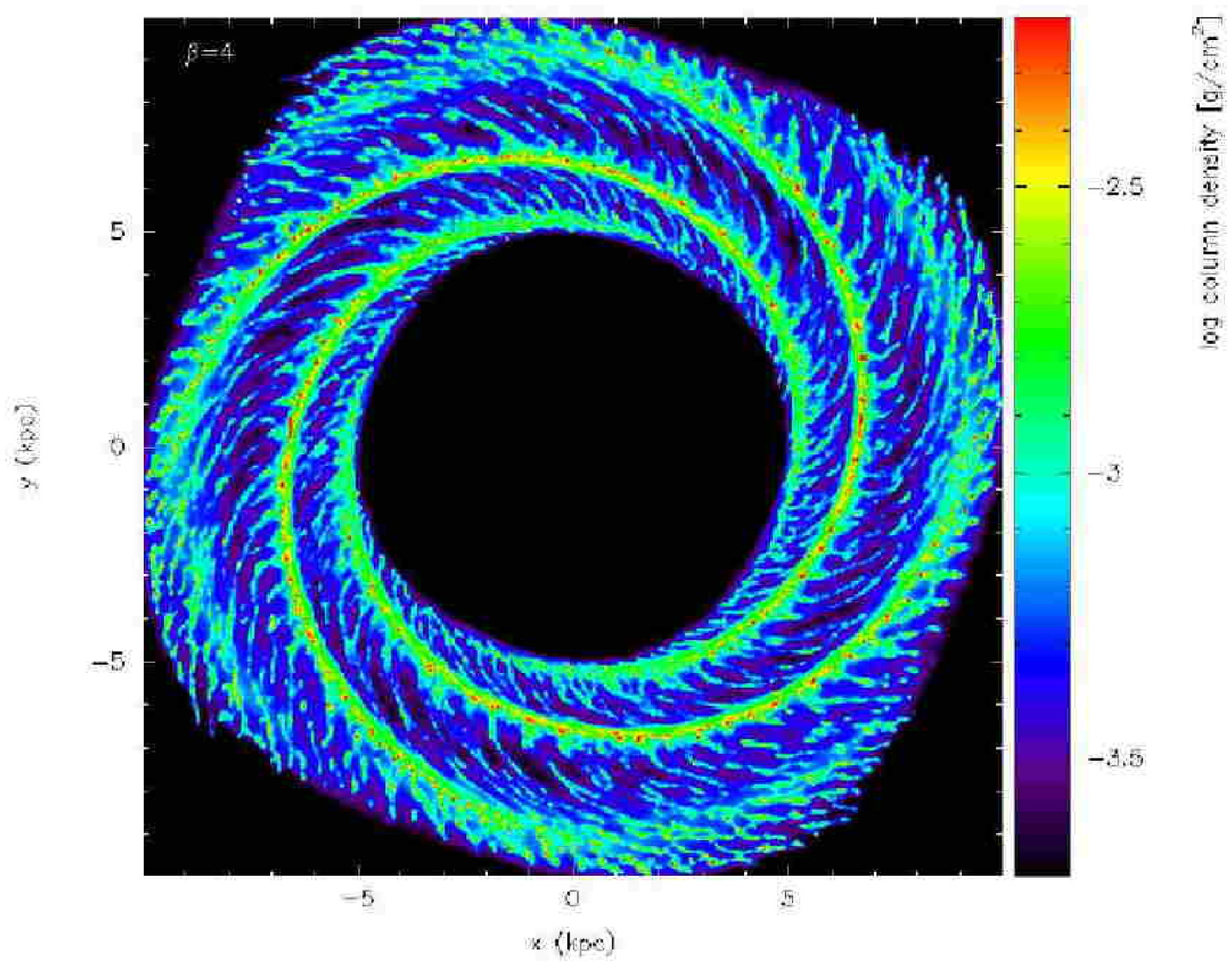} 
\includegraphics[scale=0.4]{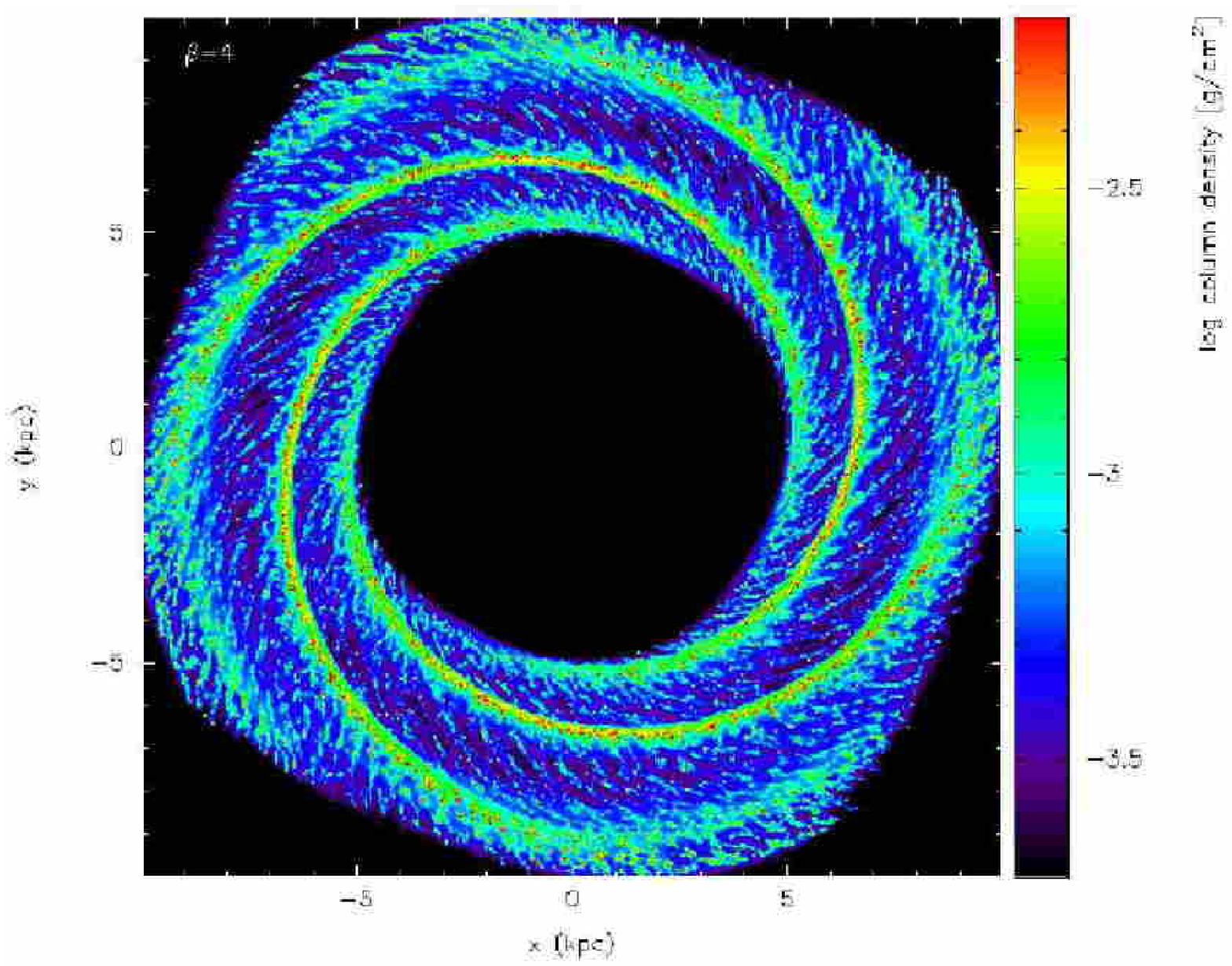}}   
\centerline{
\includegraphics[scale=0.4]{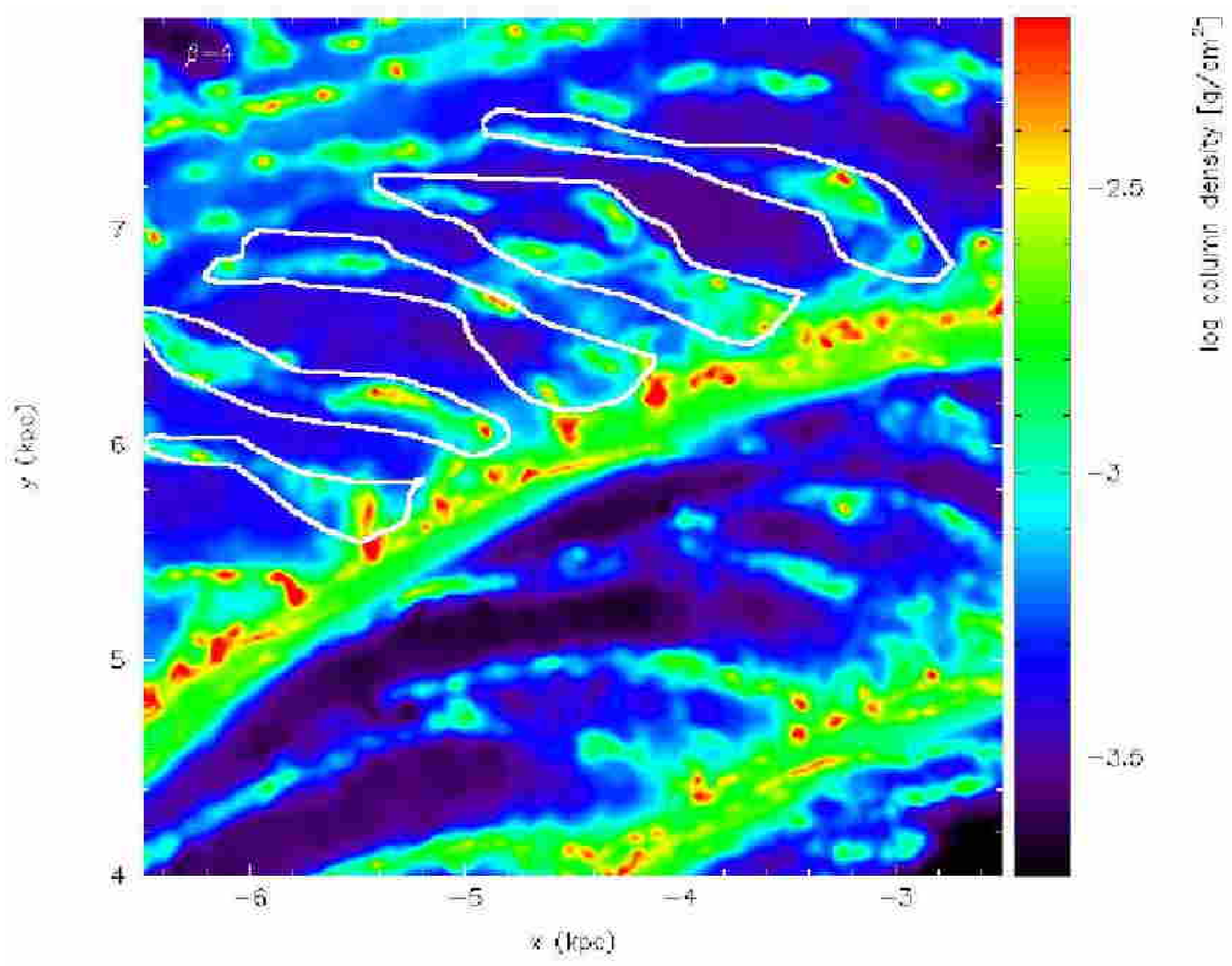}
\includegraphics[scale=0.4]{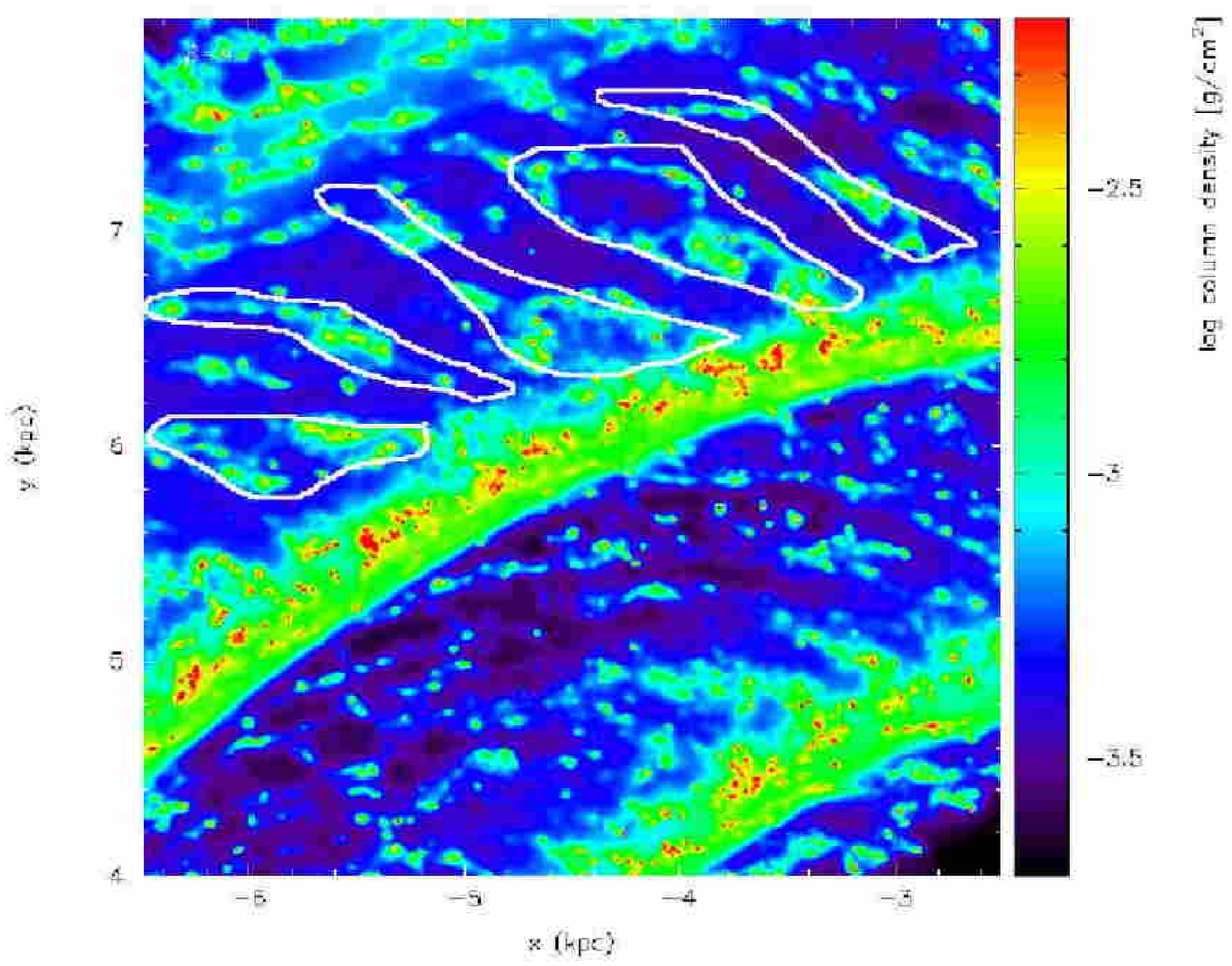}}
\caption{The column density is shown at different resoutions for the two-phase simulations with $\beta_{cold}=4$ after 250 Myr. The reolution is 1 million particles (left) and 8 million particles (right) for the whole disc (top) and a 4 x 4 kpc subsection (bottom). At higher resolution, more detailed structure emerges, and the spurs become more fragmented. However the large scale features are similar at the two different resolutions.} 
\end{figure*}
\section{Calculation of polarised synchrotron emission}
The method for computing the sample polarised synchrotron emission map shown in Section~3.5.2 is described here, following \citet*{Sokoloff1998} and \citet*{Urbanik1997}.

We adopt a reference frame such that the galaxy is exactly face-on to the observer, such that $z$ is the coordinate along the line of sight and $x,y$ are the coordinates both in the plane of the sky and also in the galactic plane. The direction (B vectors) and intensity of synchrotron polarisation are determined observationally by the Stokes Q, U and I parameters (respectively), given  by
\begin{eqnarray}
Q & = & p_{i} \int w({\bf r}) \varepsilon({\bf r}) \cos{\chi} {\rm dV}, \label{eq:stokesQ} \\
U & = & p_{i} \int w({\bf r}) \varepsilon({\bf r}) \sin{\chi} {\rm dV}, \\
I & = & p_{i} \int w({\bf r}) \varepsilon({\bf r}) {\rm dV} \label{eq:stokesI}
\end{eqnarray}
where the integration is over the cylindrical column consisting of an assumed beam profile w(x,y) along the line of sight (z),  $\epsilon$ is the local synchrotron emissivity, $p_{i}$ is the intrinsic degree of polarisation and the angle $\chi$ is related to the local magnetic field according to
\begin{equation}
\chi = \tan^{-1}(B_{y}/B_{x}) + K \lambda^{2} \int B_{z} n(r,z) {\rm dz}.
\end{equation}
The second term in the above expression introduces the effect of Faraday rotation: a wavelength-dependent ($\propto \lambda^{2}$) change of the polarisation angle dependent on the strength of the magnetic field parallel to the line of sight ($B_{z}$) and the number density of thermal electrons, $n(r,z)$ (the parameter $K$ is a constant given by $K=0.81$ rad m$^{-2}$ cm$^{3} \mu$G$ ^{-1}$ pc$^{-1}$). Since Faraday rotation is wavelength-dependent it a) becomes negligible at short wavelengths (ie. for observations made shortward of $\lambda \sim 6$cm for typical ISM magnetic fields \citep{Beck2007} and b) can be corrected for by observations made at multiple wavelengths. Thus we show our results assuming no Faraday rotation.

The synchrotron emissivity is given by:
\begin{equation}
\varepsilon({\bf r}) \propto B_{\perp}^{1+\alpha} n_{cr}(r,z)
\end{equation}
where $\alpha$ is the synchrotron spectral index ($I_{\nu}\propto \nu^{-\alpha}$), $B_{\perp} \equiv \sqrt{B_{x}^{2} + B_{y}^{2}}$ is the magnetic field in the plane of the sky and $n_{cr}(r,z)$ is the local number density of cosmic ray electrons. The intrinsic degree of polarisation, $p_{i}$ is related to the spectral index by $p_{i} = (\alpha + 1)/(\alpha + 5/3)$, which for our choice of $\alpha = 0.8$ gives $p_{i} = 0.73$.

The electron cosmic ray distribution is assumed to be of the form
\begin{equation}
n_{cr} = \exp[-r/r_{crit} - z/z_{crit}]
\end{equation}
where the distribution is assumed to fall off with galactic radius assuming $r_{crit} = 13.5$ kpc and with galactic scale height using $z_{crit} = 2.5$ kpc \citep{Gomez2004,Ferriere1998}.

In order to construct the polarisation map efficiently from our SPH simulations we use a beam profile $w$ based on the usual SPH cubic spline kernel \citep{Monaghan1992},
\begin{equation}
W(q) = \frac{1}{\pi} \left\{ \begin{array}{ll}
1 - \frac{3}{2}q^2 + \frac{3}{4}q^3, & 0 \le q < 1; \\
\frac{1}{4}(2-q)^3, & 1 \le q < 2; \\
0. & q \ge 2. \end{array} \right. \label{eq:cubicspline}
\end{equation}
where $q = \sqrt{(x^{2} + y^{2} + z^{2})}/h$. This kernel is approximately Gaussian in shape but with compact support of radius $2h$, where $h$ is the smoothing length). We perform the integration by interpolating the quantity to be integrated at a given position from nearby SPH particles, ie.
\begin{equation}
\langle A(x,y,z) \rangle = \sum_{j} \frac{m_{j}}{\rho_{j} h_{j}^{3}} A_{j} W_{ij} (x,y,z, h_{j}) \label{eq:sphsum}
\end{equation}
where $m_{j}, \rho_{j}$ and $h_{j}$ are the mass, mass density and smoothing length of particle $j$, each of which are known from the simulation. For example in the case of the Stokes Q parameter, we have $A \equiv \varepsilon(x,y,z) \cos\chi$ (similarly for the $U$ and $I$ calculations).

The use of the summation interpolant (equation \ref{eq:sphsum}) takes care of the $x,y$ dependence of the integral (having replaced it with a summation over contributing particles). The line-of-sight (z) integration in (\ref{eq:stokesQ})-(\ref{eq:stokesI}) is then performed by integrating (\ref{eq:sphsum}) analytically, giving
\begin{equation}
\int \langle A \rangle dz = \sum_{j} \frac{m_{j}}{\rho_{j} h_{j}^{3}} A_{j} \int W_{ij} (x,y,z, h) {\rm dz},
\end{equation}
such that the $z$ integration simply requires the integral of the kernel through one spatial dimension. This can be pre-computed efficiently in a table, giving, in effect
\begin{equation}
\int \langle A \rangle dz = \sum_{j} \frac{m_{j}}{\rho_{j} h_{j}^{2}} A_{j} Y_{ij} (x,y,h),
\label{eq:zsum}
\end{equation}
 where $Y$ is $W$ integrated along one spatial dimension.  To approximate a beam profile $w(x,y) \equiv Y(x,y)$ with a fixed resolution length (rather than the varying resolution length used in the SPH simulation), we set the smoothing length used in the interpolation to $h = {\rm max}(h_{j},h_{beam})$. In practice our resolution is such that $h_{beam} >> h_{j}$ (that is the simulation is much better resolved than the observations).

The net result of the procedure given above is that the construction of the polarisation map is reduced to a loop over all the particles in the simulation, where for each particle a contribution is added to all pixels in the image lying within $2h$.

The additional inclusion of Faraday rotation would add a slight complication to the calculation in that the angle $\chi$ also depends on a line-of-sight integration. This would require sorting the particles in $z$ prior to computing the sum (\ref{eq:zsum}), then computing the sum from back to front with $\chi$ stored and incrementally updated for each pixel.

The Fortran routine for performing this interpolation has been incorporated into SPLASH and is available on request. 
\bibliographystyle{mn2e}
\bibliography{Dobbs}

\begin{thebibliography}{}

\bibitem[\protect\citeauthoryear{{Balsara}, {Kim}, {Mac Low} \&
  {Mathews}}{{Balsara} et~al.}{2004}]{Balsara2004}
{Balsara} D.~S.,  {Kim} J.,  {Mac Low} M.-M.,    {Mathews} G.~J.,  2004, \apj,
  617, 339

\bibitem[\protect\citeauthoryear{{Bate}}{{Bate}}{1995}]{Bate1995}
{Bate} M.,  1995, PhD thesis, Univ.~Cambridge

\bibitem[\protect\citeauthoryear{{Beck}}{{Beck}}{2005}]{Beck2005}
{Beck} R.,  2005, in {de Gouveia dal Pino} E.~M.,  {Lugones} G.,   {Lazarian}
  A.,  eds, AIP Conf. Proc. 784: Magnetic Fields in the Universe: From
  Laboratory and Stars to Primordial Structures. {Observations of Magnetic
  Fields in Galaxies}.
pp 343--353

\bibitem[\protect\citeauthoryear{{Beck}}{{Beck}}{2007}]{Beck2007}
{Beck} R.,  2007, in EAS Publications Series {Magnetic Field Structure from
  Synchrotron Polarization}.
pp 19--36

\bibitem[\protect\citeauthoryear{{Beck}, {Brandenburg}, {Moss}, {Shukurov} \&
  {Sokoloff}}{{Beck} et~al.}{1996}]{Beck1996}
{Beck} R.,  {Brandenburg} A.,  {Moss} D.,  {Shukurov} A.,    {Sokoloff} D.,
  1996, \araa, 34, 155

\bibitem[\protect\citeauthoryear{{Beck} \& {Hoernes}}{{Beck} \&
  {Hoernes}}{1996}]{BeckNat1996}
{Beck} R.,  {Hoernes} P.,  1996, \nat, 379, 47

\bibitem[\protect\citeauthoryear{{Benz}, {Cameron}, {Press} \& {Bowers}}{{Benz}
  et~al.}{1990}]{Benz1990}
{Benz} W.,  {Cameron} A.~G.~W.,  {Press} W.~H.,    {Bowers} R.~L.,  1990, \apj,
  348, 647

\bibitem[\protect\citeauthoryear{{Binney} \& {Tremaine}}{{Binney} \&
  {Tremaine}}{1987}]{Binney}
{Binney} J.,  {Tremaine} S.,  1987, {Galactic dynamics}.
Princeton, NJ, Princeton University Press, 1987, 747 p.

\bibitem[\protect\citeauthoryear{{Bonnell}, {Dobbs}, {Robitaille} \&
  {Pringle}}{{Bonnell} et~al.}{2006}]{Bonnell2006}
{Bonnell} I.~A.,  {Dobbs} C.~L.,  {Robitaille} T.~P.,    {Pringle} J.~E.,
  2006, \mnras, 365, 37

\bibitem[\protect\citeauthoryear{{Chakrabarti}, {Laughlin} \&
  {Shu}}{{Chakrabarti} et~al.}{2003}]{Chak2003}
{Chakrabarti} S.,  {Laughlin} G.,    {Shu} F.~H.,  2003, \apj, 596, 220

\bibitem[\protect\citeauthoryear{{Cox}}{{Cox}}{2005}]{Cox2005}
{Cox} D.~P.,  2005, \araa, 43, 337

\bibitem[\protect\citeauthoryear{{Cox} \& {G{\' o}mez}}{{Cox} \& {G{\'
  o}mez}}{2002}]{Cox2002}
{Cox} D.~P.,  {G{\' o}mez} G.~C.,  2002, \apjs, 142, 261

\bibitem[\protect\citeauthoryear{{Dobbs}}{{Dobbs}}{2007}]{Thesis}
{Dobbs} C.~L.,  2007, PhD thesis, Univ.~of St Andrews

\bibitem[\protect\citeauthoryear{{Dobbs} \& {Bonnell}}{{Dobbs} \&
  {Bonnell}}{2006}]{Dobbs2006}
{Dobbs} C.~L.,  {Bonnell} I.~A.,  2006, \mnras, 367, 873

\bibitem[\protect\citeauthoryear{{Dobbs} \& {Bonnell}}{{Dobbs} \&
  {Bonnell}}{2007a}]{Dobbs2007a}
{Dobbs} C.~L.,  {Bonnell} I.~A.,  2007a, \mnras, 374, 1115

\bibitem[\protect\citeauthoryear{{Dobbs} \& {Bonnell}}{{Dobbs} \&
  {Bonnell}}{2007b}]{Dobbs2007}
{Dobbs} C.~L.,  {Bonnell} I.~A.,  2007b, \mnras, 376, 1747

\bibitem[\protect\citeauthoryear{{Dobbs}, {Bonnell} \& {Pringle}}{{Dobbs}
  et~al.}{2006}]{DBP2006}
{Dobbs} C.~L.,  {Bonnell} I.~A.,    {Pringle} J.~E.,  2006, \mnras, 371, 1663

\bibitem[\protect\citeauthoryear{{Elmegreen}}{{Elmegreen}}{1982}]{Elmegreen198%
2}
{Elmegreen} B.~G.,  1982, \apj, 253, 655

\bibitem[\protect\citeauthoryear{{Elmegreen}}{{Elmegreen}}{1987}]{ElmegreenMJI%
1987}
{Elmegreen} B.~G.,  1987, \apj, 312, 626

\bibitem[\protect\citeauthoryear{{Ferriere}}{{Ferriere}}{1998}]{Ferriere1998}
{Ferriere} K.,  1998, \apj, 497, 759

\bibitem[\protect\citeauthoryear{{Gibson}, {Taylor}, {Stil}, {Brunt}, {Kavars}
  \& {Dickey}}{{Gibson} et~al.}{2006}]{Gibson2006}
{Gibson} S.~J.,  {Taylor} A.~R.,  {Stil} J.~M.,  {Brunt} C.~M.,  {Kavars}
  D.~W.,    {Dickey} J.~M.,  2006, in IAU Symposium {Cold HI in Turbulent
  Eddies and Galactic Spiral Shocks}

\bibitem[\protect\citeauthoryear{{Gittins} \& {Clarke}}{{Gittins} \&
  {Clarke}}{2004}]{Gittins2004}
{Gittins} D.~M.,  {Clarke} C.~J.,  2004, \mnras, 349, 909

\bibitem[\protect\citeauthoryear{{G{\'o}mez} \& {Cox}}{{G{\'o}mez} \&
  {Cox}}{2002}]{Gomez2002}
{G{\'o}mez} G.~C.,  {Cox} D.~P.,  2002, \apj, 580, 235

\bibitem[\protect\citeauthoryear{{G{\'o}mez} \& {Cox}}{{G{\'o}mez} \&
  {Cox}}{2004}]{Gomez2004}
{G{\'o}mez} G.~C.,  {Cox} D.~P.,  2004, \apj, 615, 744

\bibitem[\protect\citeauthoryear{{Han}, {Manchester}, {Lyne}, {Qiao} \& {van
  Straten}}{{Han} et~al.}{2006}]{Han2006}
{Han} J.~L.,  {Manchester} R.~N.,  {Lyne} A.~G.,  {Qiao} G.~J.,    {van
  Straten} W.,  2006, \apj, 642, 868

\bibitem[\protect\citeauthoryear{{Heiles} \& {Crutcher}}{{Heiles} \&
  {Crutcher}}{2005}]{Heiles2005}
{Heiles} C.,  {Crutcher} R.,  2005, in {Wielebinski} R.,  {Beck} R.,  eds, LNP
  Vol. 664: Cosmic Magnetic Fields {Magnetic Fields in Diffuse HI and Molecular
  Clouds}.
pp 137--+

\bibitem[\protect\citeauthoryear{{Heiles} \& {Troland}}{{Heiles} \&
  {Troland}}{2003}]{Heiles2003}
{Heiles} C.,  {Troland} T.~H.,  2003, \apj, 586, 1067

\bibitem[\protect\citeauthoryear{{Heiles} \& {Troland}}{{Heiles} \&
  {Troland}}{2005}]{HT2005}
{Heiles} C.,  {Troland} T.~H.,  2005, \apj, 624, 773

\bibitem[\protect\citeauthoryear{{Kim}, {Kim} \& {Ostriker}}{{Kim}
  et~al.}{2006}]{KKO2006}
{Kim} C.-G.,  {Kim} W.-T.,    {Ostriker} E.~C.,  2006, \apjl, 649, L13

\bibitem[\protect\citeauthoryear{{Kim}, {Ryu} \& {Jones}}{{Kim}
  et~al.}{2001}]{OtherKim2001}
{Kim} J.,  {Ryu} D.,    {Jones} T.~W.,  2001, \apj, 557, 464

\bibitem[\protect\citeauthoryear{{Kim} \& {Ostriker}}{{Kim} \&
  {Ostriker}}{2002}]{Kim2002}
{Kim} W.,  {Ostriker} E.~C.,  2002, \apj, 570, 132

\bibitem[\protect\citeauthoryear{{Kim}, {Ostriker} \& {Stone}}{{Kim}
  et~al.}{2002}]{KOS2002}
{Kim} W.,  {Ostriker} E.~C.,    {Stone} J.~M.,  2002, \apj, 581, 1080

\bibitem[\protect\citeauthoryear{{Kim} \& {Ostriker}}{{Kim} \&
  {Ostriker}}{2001}]{Kim2001}
{Kim} W.-T.,  {Ostriker} E.~C.,  2001, \apj, 559, 70

\bibitem[\protect\citeauthoryear{{Kim} \& {Ostriker}}{{Kim} \&
  {Ostriker}}{2006}]{Kim2006}
{Kim} W.-T.,  {Ostriker} E.~C.,  2006, \apj, 646, 213

\bibitem[\protect\citeauthoryear{{Kim}, {Ostriker} \& {Stone}}{{Kim}
  et~al.}{2003}]{Kim2003}
{Kim} W.-T.,  {Ostriker} E.~C.,    {Stone} J.~M.,  2003, \apj, 599, 1157

\bibitem[\protect\citeauthoryear{{Kosi{\'n}ski} \& {Hanasz}}{{Kosi{\'n}ski} \&
  {Hanasz}}{2007}]{Kosinski2007}
{Kosi{\'n}ski} R.,  {Hanasz} M.,  2007, \mnras, 376, 861

\bibitem[\protect\citeauthoryear{{Laing}}{{Laing}}{1980}]{Laing1980}
{Laing} R.~A.,  1980, \mnras, 193, 439

\bibitem[\protect\citeauthoryear{{Li}, {Griffin}, {Krejny}, {Novak},
  {Loewenstein}, {Newcomb}, {Calisse} \& {Chuss}}{{Li} et~al.}{2006}]{LiG2006}
{Li} H.,  {Griffin} G.~S.,  {Krejny} M.,  {Novak} G.,  {Loewenstein} R.~F.,
  {Newcomb} M.~G.,  {Calisse} P.~G.,    {Chuss} D.~T.,  2006, \apj, 648, 340

\bibitem[\protect\citeauthoryear{{McKee}}{{McKee}}{1990}]{McKee1990}
{McKee} C.~F.,  1990, in {Blitz} L.,  ed., The Evolution of the Interstellar
  Medium Vol.~12 of Astronomical Society of the Pacific Conference Series, {The
  three phase model of the interstellar medium - Where does it stand now?}.
pp 3--29

\bibitem[\protect\citeauthoryear{{Monaghan}}{{Monaghan}}{1992}]{Monaghan1992}
{Monaghan} J.~J.,  1992, \araa, 30, 543

\bibitem[\protect\citeauthoryear{{Monaghan}}{{Monaghan}}{1997}]{Monaghan1997}
{Monaghan} J.~J.,  1997, Journal of Computational Physics, 136, 298

\bibitem[\protect\citeauthoryear{{Neininger}}{{Neininger}}{1992}]{Nein1992}
{Neininger} N.,  1992, \aap, 263, 30

\bibitem[\protect\citeauthoryear{{Nishikori}, {Machida} \&
  {Matsumoto}}{{Nishikori} et~al.}{2006}]{Nish2006}
{Nishikori} H.,  {Machida} M.,    {Matsumoto} R.,  2006, \apj, 641, 862

\bibitem[\protect\citeauthoryear{{Panesar} \& {Nelson}}{{Panesar} \&
  {Nelson}}{1992}]{Panesar1992}
{Panesar} J.~S.,  {Nelson} A.~H.,  1992, \aap, 264, 77

\bibitem[\protect\citeauthoryear{{Parker}}{{Parker}}{1966}]{Parker1966}
{Parker} E.~N.,  1966, \apj, 145, 811

\bibitem[\protect\citeauthoryear{{Parker}}{{Parker}}{1971a}]{Parker1971}
{Parker} E.~N.,  1971a, \apj, 163, 255

\bibitem[\protect\citeauthoryear{{Parker}}{{Parker}}{1971b}]{P1971}
{Parker} E.~N.,  1971b, \apj, 163, 279

\bibitem[\protect\citeauthoryear{{Patrikeev}, {Fletcher}, {Stepanov}, {Beck},
  {Berkhuijsen}, {Frick} \& {Horellou}}{{Patrikeev}
  et~al.}{2006}]{Patrikeev2006}
{Patrikeev} I.,  {Fletcher} A.,  {Stepanov} R.,  {Beck} R.,  {Berkhuijsen}
  E.~M.,  {Frick} P.,    {Horellou} C.,  2006, \aap, 458, 441

\bibitem[\protect\citeauthoryear{{Piontek} \& {Ostriker}}{{Piontek} \&
  {Ostriker}}{2005}]{Piontek2005}
{Piontek} R.~A.,  {Ostriker} E.~C.,  2005, \apj, 629, 849

\bibitem[\protect\citeauthoryear{{Price}}{{Price}}{2007}]{Price2007}
{Price} D.~J.,  2007, astro-ph/07090832

\bibitem[\protect\citeauthoryear{{Price} \& {Bate}}{{Price} \&
  {Bate}}{2007}]{PB2007}
{Price} D.~J.,  {Bate} M.~R.,  2007, \mnras, 377, 77

\bibitem[\protect\citeauthoryear{{Price} \& {Monaghan}}{{Price} \&
  {Monaghan}}{2004a}]{Price2004b}
{Price} D.~J.,  {Monaghan} J.~J.,  2004a, \mnras, 348, 123

\bibitem[\protect\citeauthoryear{{Price} \& {Monaghan}}{{Price} \&
  {Monaghan}}{2004b}]{Price2004}
{Price} D.~J.,  {Monaghan} J.~J.,  2004b, \mnras, 348, 139

\bibitem[\protect\citeauthoryear{{Price} \& {Monaghan}}{{Price} \&
  {Monaghan}}{2005}]{Price2005}
{Price} D.~J.,  {Monaghan} J.~J.,  2005, \mnras, 364, 384

\bibitem[\protect\citeauthoryear{{Price} \& {Monaghan}}{{Price} \&
  {Monaghan}}{2007}]{PM2007}
{Price} D.~J.,  {Monaghan} J.~J.,  2007, \mnras, 374, 1347

\bibitem[\protect\citeauthoryear{{Priest}}{{Priest}}{1982}]{Priest1982}
{Priest} E.~R.,  1982, {Solar magneto-hydrodynamics}.
Dordrecht, Holland ; Boston : D.~Reidel Pub.~Co.~; Hingham,

\bibitem[\protect\citeauthoryear{{Roberts}}{{Roberts}}{1969}]{Roberts1969}
{Roberts} W.~W.,  1969, \apj, 158, 123

\bibitem[\protect\citeauthoryear{{Roberts} Jr. \& {Yuan}}{{Roberts} \&
  {Yuan}}{1970}]{Roberts1970}
{Roberts} Jr. W.~W.,  {Yuan} C.,  1970, \apj, 161, 887

\bibitem[\protect\citeauthoryear{{Rohde} \& {Elstner}}{{Rohde} \&
  {Elstner}}{1998}]{Rohde1998}
{Rohde} R.,  {Elstner} D.,  1998, \aap, 333, 27

\bibitem[\protect\citeauthoryear{{Shetty} \& {Ostriker}}{{Shetty} \&
  {Ostriker}}{2006}]{Shetty2006}
{Shetty} R.,  {Ostriker} E.~C.,  2006, \apj, 647, 997

\bibitem[\protect\citeauthoryear{{Shu}, {Adams} \& {Lizano}}{{Shu}
  et~al.}{1987}]{Shu1987}
{Shu} F.~H.,  {Adams} F.~C.,    {Lizano} S.,  1987, \araa, 25, 23

\bibitem[\protect\citeauthoryear{{Shu}, {Milione}, {Gebel}, {Yuan}, {Goldsmith}
  \& {Roberts}}{{Shu} et~al.}{1972}]{Shu1972}
{Shu} F.~H.,  {Milione} V.,  {Gebel} W.,  {Yuan} C.,  {Goldsmith} D.~W.,
  {Roberts} W.~W.,  1972, \apj, 173, 557

\bibitem[\protect\citeauthoryear{{Slyz}, {Kranz} \& {Rix}}{{Slyz}
  et~al.}{2003}]{Slyz2003}
{Slyz} A.~D.,  {Kranz} T.,    {Rix} H.-W.,  2003, \mnras, 346, 1162

\bibitem[\protect\citeauthoryear{{Sokoloff}, {Bykov}, {Shukurov},
  {Berkhuijsen}, {Beck} \& {Poezd}}{{Sokoloff} et~al.}{1998}]{Sokoloff1998}
{Sokoloff} D.~D.,  {Bykov} A.~A.,  {Shukurov} A.,  {Berkhuijsen} E.~M.,  {Beck}
  R.,    {Poezd} A.~D.,  1998, \mnras, 299, 189

\bibitem[\protect\citeauthoryear{{Stern}}{{Stern}}{1970}]{Stern1970}
{Stern} D.~P.,  1970, American Journal of Physics, 38, 494

\bibitem[\protect\citeauthoryear{{Tanaka}, {Wada}, {Machida}, {Matsumoto} \&
  {Miyaji}}{{Tanaka} et~al.}{2005}]{Tanaka2005}
{Tanaka} M.,  {Wada} K.,  {Machida} M.,  {Matsumoto} R.,    {Miyaji} S.,  2005,
  in {de Gouveia dal Pino} E.~M.,  {Lugones} G.,   {Lazarian} A.,  eds, AIP
  Conf. Proc. 784: Magnetic Fields in the Universe: From Laboratory and Stars
  to Primordial Structures. {Magnetohydrodynamic Simulations of the Wiggle
  Instability in Spiral Galaxies}.
pp 792--797

\bibitem[\protect\citeauthoryear{{Urbanik}, {Elstner} \& {Beck}}{{Urbanik}
  et~al.}{1997}]{Urbanik1997}
{Urbanik} M.,  {Elstner} D.,    {Beck} R.,  1997, \aap, 326, 465

\bibitem[\protect\citeauthoryear{{Vallee}}{{Vallee}}{1995}]{Vallee1995}
{Vallee} J.~P.,  1995, \apss, 234, 1

\bibitem[\protect\citeauthoryear{{Wada} \& {Koda}}{{Wada} \&
  {Koda}}{2004}]{Wada2004}
{Wada} K.,  {Koda} J.,  2004, \mnras, 349, 270

\bibitem[\protect\citeauthoryear{{Wolfire}, {McKee}, {Hollenbach} \&
  {Tielens}}{{Wolfire} et~al.}{2003}]{Wolfire2003}
{Wolfire} M.~G.,  {McKee} C.~F.,  {Hollenbach} D.,    {Tielens} A.~G.~G.~M.,
  2003, \apj, 587, 278

\bibitem[\protect\citeauthoryear{{Woodward}}{{Woodward}}{1976}]{Woodward1976}
{Woodward} P.~R.,  1976, \apj, 207, 484

\bibitem[\protect\citeauthoryear{{Zweibel} \& {Kulsrud}}{{Zweibel} \&
  {Kulsrud}}{1975}]{Zweibel1975}
{Zweibel} E.~G.,  {Kulsrud} R.~M.,  1975, \apj, 201, 63

\end{thebibliography}

\bsp

\label{lastpage}

\end{document}